\documentclass[12pt,hyper]{JHEP3}  
\pdfoutput=1
\usepackage{amsmath}
\usepackage{amssymb}
\usepackage{graphicx}
\usepackage{multirow}
\usepackage{epsfig}
\usepackage{rotating}
\usepackage{longtable} 
\usepackage{url}
\usepackage{pdflscape}
\usepackage[utf8]{inputenc}
\usepackage{csquotes}
\usepackage{placeins}

\newcommand{\cc}{\c c}
\newcommand{\vectorc}{$g_{2V}(f)$}
\newcommand{\axialc}{$g_{2A}(f)$}

\title{LHC Constraints on 3-3-1 Models}
\author{Camilo Salazar$^{1}$, Richard H.~Benavides$^{3}$, William A.~Ponce$^{1}$ and  Eduardo Rojas$^{1,2}$
\\
$^{1}$Instituto  de F\'isica,  \\ 
Universidad de Antioquia, Calle 70 No. 52-21, Medell\'in, Colombia \\
$^{2}$Laboratorio de F\'isica 
Te\'orica e Computa\cc\~ao Cient\'ifica, 
Universidade Cruzeiro do Sul, 01506-000, S\~ao Paulo, Brazil. \\
$^{3}$Facultad de Ciencias Exactas y Aplicadas, Instituto Tecnológico Metropolitano, Calle 73 No 76 A - 354 , Vía el Volador, Medellín, Colombia. 
\\
\email{camilo@gfif.udea.edu.co} \\
\email{richardbenavides@itm.edu.co} \\
\email{wponce@fisica.udea.edu.co} \\
\email{eduardo.rojas@cruzeirodosul.edu.br}
}

\newcommand{\be}{\begin{equation}}
\newcommand{\ee}{\end{equation}}
\newcommand{\bea}{\begin{eqnarray}}
\newcommand{\eea}{\end{eqnarray}}
\newcommand{\ba}{\begin{array}}
\newcommand{\ea}{\end{array}}
\newcommand{\nn}{\nonumber}
 
\newcommand{\ie}{{\emph i.e.,\ }}

\newcommand{\sbx}{\scalebox}
\newcommand{\M}{M_{l^+l^-}}


\date{}

\abstract{The ATLAS detector data on di-lepton production is used 
in order to impose constraints on $Z^\prime$ boson masses associated with a variety of  3-3-1 and $E_6$ motivated  $Z'$ models. 
Lower mass bounds for the  different models are established at $95\%$ confidence level. 
Our numerical analysis is extrapolated up to 14~TeV, and further to 30~TeV and 100~TeV, for a broad range of  luminosities. 
Some of our results can be compared with the ATLAS published bounds, being, for those cases, in fairly good agreement.
We also  report the vector and axial charges for all the   3-3-1-motivated $Z'$ models 
without exotic electric charges for leptons,
known in the literature. To the best of our knowledge  most of this charges were not reported before.}
\preprint{}
\begin{document}
\section{Introduction}
\label{sec:intro}
The existence of new neutral vector bosons $Z'$ beyond the one associated with the $SU(3)_c\otimes SU(2)_L\otimes U(1)_Y$ 
local gauge group of the Standard Model (SM)\footnote{For an excellent compendium of the SM, see Ref.~\cite{Donoghue:1992dd}.}, 
is a clear prediction of new physics, related to extra $U(1)$ factors 
appearing  in every regular chain of the  breaking of  larger gauge groups down to the SM one~\cite{Langacker:2008yv}.

A systematic study of additional $U(1)$ symmetries is possible just by restricting to the study of the lowest dimensional 
representations of larger gauge groups and their branching rules~\cite{Slansky:1981yr}. As it is well known, a family
non universal $Z'$ coupling leads to Flavor-Changing Neutral Currents~(FCNC), and possibly to new CP-violating effects~\cite{Langacker:2000ju}. 
To avoid these inconveniences, some of the first models with physics beyond the SM incorporate the 
assumption of family  universality, condition quite restrictive to such an extent that, it is not possible to construct  
a minimal extension of the SM just by adding a $U(1)$ factor to the SM local gauge group, without the introduction of new 
fermion fields~\cite{Langacker:2008yv,Appelquist:2002mw,Carena:2004xs}; that is, it is not possible a $Z'$  interaction 
just with the current content of the  particles in the SM.

The requirement of universality for the $U(1)$ charges and, in consequence anomaly cancellation in every family, 
leads in a natural way into $E_6$ subgroups in most of the cases. As a gauge group, $E_6$ is the only exceptional group with complex representations 
that  is anomaly free in 
all  its representations~\cite{Gursey:1975ki}.   Some  $E_6$ subgroups, such as the  original unification groups  $SU(5)$,  
$SO(10)$, and the Left-Right symmetric models $SU(4)\times SU(2)_L\times SU(2)_R$ with their corresponding  supersymmetric 
realizations, are between the most widely  known extensions of the SM. 
For a classification of $U(1)'$ symmetries contained in $E_6$ see references~\cite{Robinett:1982tq, Erler:2011ud}. 

Early in the nineties, some work pointed out to the conclusion that universality must not be taken for granted for models with physics beyond the SM. 
In particular, under some suitable assumptions,
many non universal models were able to evade the FCNC constraints. Following this trend of ideas, 
the $SU(3)_c\otimes SU(3)_L\otimes U(1)_x$ models (3-3-1 for short) were  proposed  by allowing anomaly cancellation between fermions 
in different families \cite{Singer:1980sw,Pisano:1991ee,Frampton:1992wt,Montero:1992jk,Foot:1992rh,Foot:1994ym,Ozer:1995xi,Ponce:2001jn,Ponce:2002sg}.

For the most popular 3-3-1 models~\cite{Pisano:1991ee,Foot:1992rh,Ozer:1995xi}, three families is the simplest possible choice of matter content
in order to have anomaly cancellation. So, one of the most  appealing features 
of those models is to provide explanation for the family replication problem (also known as the generation number problem), which 
is a long standing issue in particle physics; furthermore, they provide some indications
of why the top family is the heaviest one~\cite{Ozer:1996jc}. Also, 3-3-1 models are  among the most interesting new physics scenarios
with new sources of CP and flavor
violation~\cite{Montero:2005yb}, making them the most suitable ones for  flavor 
studies~\cite{Buras:2004uu,Buras:2012dp,Buras:2013ooa,Buras:2014yna,Promberger:2007py,Cabarcas:2011hb,Machado:2013jca,Martinez:2014lta}.

The first 3-3-1 model for three families was sketched originally in Ref.~\cite{Singer:1980sw}, where references 
to previous $SU(3)\otimes U(1)$ models for  one and two families can be found. 
Then, in Refs.~\cite{Pisano:1991ee,Frampton:1992wt} the so called minimal version of the model was introduced, minimal in the sense that 
it does not contain lepton fields beyond the ones present in the SM. Next, came the 3-3-1 family model with right handed neutrinos, 
rediscovered in Refs.~\cite{Montero:1992jk,Foot:1992rh,Foot:1994ym} 
(the first 3-3-1 family model with right handed neutrinos was introduced in Ref.~\cite{Singer:1980sw}). 
The three family model with exotic electrons was introduced in the literature in Ref.~\cite{Ozer:1995xi}, 
a classification of 3-3-1 models without exotic electric charges was done in Refs.~\cite{Ponce:2001jn}; 
and finally, the so called economical 3-3-1 model appeared in Ref.~\cite{Ponce:2002sg}.

Since the gauge group for the 3-3-1 models is not simple, neither semi-simple, there is not a neat prediction of the electroweak mixing angle, 
neither there is an explanation for the quantization of the electric charge using only the 
cancellation of anomalies (the quantum constraints)~\footnote{In grand unified theories with simple gauge groups,  
the electric charge is quantized because the charge operator is a linear combination of generators of the unifying group.}; 
but, as in the SM, the inclusion of the classical constraints leads in a simple way to the quantization of the electric charge~\cite{deSousaPires:1998jc}, 
conclusion linked to the generation number problem in Ref.~\cite{VanDong:2005ux}. 
As a last remark, it has been shown that the most general Yukawa couplings in some 3-3-1 models, 
include in a natural way a Peccei-Quinn type symmetry that can be extended to the entire Lagrangian in a very elegant way~\cite{Pal:1994ba}, 
and by using appropriate extra fields, the resulting axion can be made invisible.

In the eventual discovery of a new neutral vector boson, it will be  important the experimental determination 
of its coupling to the standard model fermions. However, the discrimination between the possible $Z'$ models
could be  challenging at the LHC, owing  to the reduced number of  high resolution channels in hadron colliders.
So, in order to carry out the statistical analysis, it is necessary to combine the LHC data with 
electroweak precision data. For 3-3-1 models, the most important constraints come 
from the flavor changing neutral currents (FCNC); in consequence, it is important to establish the models 
for which the LHC  and/or the FCNC constraints  are dominant; that is, which
kind of constraints exclude a wider region in the parameter space. It is also important to set the range of parameters and models for which 
the LHC and the FCNC constraints are comparable to each other; in such a case,  
it is convenient to combine both.

In order to set the present 95$\%$ confidence level (CL) limits and the  projected ones, we follow closely the
CDF methods explained in Ref.~\cite{Aaltonen:2007ps}. For the  exact  expression  
of the $\chi^2$ function, the theoretical formulas of the SM expected values, 
and the statistical analysis, we follow the work of the authors in Refs.~\cite{Erler:2011ud,Erler:2011iw}.
As an improvement, we update the program used in~\cite{Erler:2011ud} with the set
CTQ10 of parton distribution functions~\cite{Lai:2010vv} which allow us to reach 
higher energies than previous releases.

In this paper we present the $Z'$ charges for all the 3-3-1 models  Without Exotic Electric Charges for leptons,  known in the literature, most of them new results. Then, 
using the  recent dilepton data  reported by ATLAS in reference~\cite{Aad:2014cka} we calculate the lower bounds 
 for  $M_{Z'}$ at 95$\%$ CL, and project also at 95$\%$ CL   for the LHC and VLHC\footnote{VLHC stands for Very Large Hadron Collider 
 that would accelerate protons to energies of about 100 TeV~\cite{Godfrey:2013eta,Rizzo:2014xma}.} forthcoming energies and luminosities.
 
The paper is organized as follows: in section \ref{sec:models} we review the different 3-3-1 models present 
in the literature; in section \ref{sec:limits} we derive the present 95$\%$ CL limits and the projected ones 
on the  $Z'$ mass  for typical LHC energies and luminosities. 
The  section~\ref{sec:conclusions} summarizes  our conclusions.
Technical appendixes at the end present 
the differential cross-section formulas used in the analysis and 
the charges of the SM fermions for the different 3-3-1 models  Without exotic electric charges for leptons, in the literature .

\section{\label{sec:models}$SU(3)_c\times SU(3)_L\times U(1)_x$ Models}
The different models based on a 3-3-1 gauge symmetry are classified according to the electric charge operator which is given by
\begin{equation}\label{qem}
Q=a \lambda_3+ \frac{1}{\sqrt{3}}b\lambda_8 +xI_3,
\end{equation}
where $\lambda_\alpha,\;\alpha=1,2,\dots ,8$ are the Gell-Mann matrices for $SU(3)_L$ normalized as 
Tr$(\lambda_\alpha\lambda_\beta)=2\delta_{\alpha\beta}$ and $I_3=Dg(1,1,1)$ 
is the diagonal $3\times 3$ unit matrix. $a=1/2$ if one assumes that the isospin $SU(2)_L$ of the SM is entirely 
embedded in $SU(3)_L$ and $b$ is a free parameter which defines the different possible models. The $x$ values must be obtained by anomaly cancellation.

The covariant derivative for the electroweak sector is given now by:
\begin{equation}\label{cde}
 D_\mu=\partial_\mu-i\frac{g}{2}\sum_{\alpha=1}^8\lambda_\alpha A_\mu^\alpha-ig_1xX_\mu I_3,
\end{equation}
where $A_\mu^\alpha$ and $X_\mu$ are the gauge fields 
of $SU(3)_L$ and $U(1)_x$ respectively, and $g$ and $g_1$ are the coupling constants of the same gauge structures.

$x=0$ for $A_\mu^\alpha$, the 8 gauge fields of $SU(3)_L$, and thus Eq.~(\ref{qem}) implies:

\begin{equation}\label{gfi}
\sum_\alpha\lambda_\alpha A^\alpha_\mu=\sqrt{2}\left(
\begin{array}{ccc}
D^0_{1\mu} & W^+_\mu & K_\mu^{(b+1/2)} \\
W^-_\mu & D^0_{2\mu} & K_\mu^{(b-1/2)} \\
K_\mu^{-(b+1/2)} & K_\mu^{-(b-1/2)} & D^0_{3\mu} 
\end{array}
\right),
\end{equation}
where 
$W_\mu^\pm=(A_\mu^1\pm iA_\mu^2)/\sqrt{2},\;\; K_\mu^{\pm(b+1/2)}=(A_\mu^4\pm iA_\mu^5)/\sqrt{2},\;\; K_\mu^{\pm(b-1/2)}
=(A_\mu^6\pm iA_\mu^7)/\sqrt{2},\;\; D^0_{1\mu}=A_\mu^3/\sqrt{2}+A_\mu^8/\sqrt{6},\; 
D^0_{2\mu}=-A_\mu^3/\sqrt{2}+A_\mu^8/\sqrt{6},$ and 
$D^0_{3\mu}= -2A_\mu^8/\sqrt{6}.$  The upper index on the gauge bosons stand for the 
electric charge of the particles, some of them being functions of the $b$ parameter.

In this paper we  consider all the 3-3-1 models which do not 
include leptons with exotic electric charges; they correspond
to the $b$ parameter in equation (2.1) equal only to $\pm 1/2$ and 3/2. 
Recently, $b$ has been used as a free parameter for doing FCNC phenomenology 
in the context if 3-3-1 model \cite{Diaz:2004fs,CarcamoHernandez:2005ka,Buras:2012dp,Buras:2014yna}; 
in some of those papers~\cite{Buras:2012dp}, fermion and gauge bosons structures have been constructed for arbitrary $b$ values, 
in particular, field structures for $b=\pm 1/2,\pm 1,3/2$
are considered (for  $b\pm 1$, gauge and lepton 
fields with half integer electric charges are present).

\subsection{\label{sec:sec11}The Minimal Model}

In Refs.~\cite{Pisano:1991ee,Frampton:1992wt,Ozer:1996jc,Pleitez:1993gc,Ng:1992st,Epele:1995vv}
it was shown that, for $b=3/2$ in Eq.~(\ref{qem}), the following fermion structure is free of all the gauge anomalies: 

$\psi_{lL}^T= (l^-,\nu_l^0,l^+)_L\sim (1,3^*,0)$,  $Q_{iL}^T=(u_i,d_i,X_i)_L\sim (3,3,-1/3)$, 
$Q_{3L}^T=(d_3,u_3,Y)\sim (3,3^*,2/3)$,
where $l=e,\mu,\tau$ is a family lepton index, $i=1,2$ for the first two quark families, 
and the numbers after the similarity sign means 3-3-1 representations. The right handed fields are 
$u_{aL}^c\sim (3^*,1,-2/3),\; d_{aL}^c\sim (3^*,1,1/3),\; X_{iL}^c\sim(3^*,1,4/3)$ and $Y_L^c\sim (3^*,1,-5/3)$, 
where $a=1,2,3$ is the quark family index and there are three exotic quarks, two with electric charge $-4/3\; (X_i)$ 
and other with electric charge 5/3 $(Y)$. This version is called {\it minimal} in the literature, because
its lepton content is just the one present in the SM.

\subsection{\label{sec:sec12}3-3-1 Models Without Exotic Electric Charges}

If one wishes to avoid exotic electric charges as the ones present for the new quarks in the minimal model, one must choose $b=\pm 1/2$, in Eq.~(\ref{qem}).
Following \cite{Ponce:2002sg} we start with the following six sets of fermions which are closed in the sense that they contain
the antiparticles of the charged particles:
\begin{itemize}
\item $S_1=[(\nu^0_\alpha,\alpha^-,E_\alpha^-);\alpha^+;E_\alpha^+]_L$ with quantum numbers $(1,3,-2/3);(1,1,1)$ and $(1,1,1)$ respectively.
\item $S_2=[(\alpha^-,\nu_\alpha,N_\alpha^0);\alpha^+]_L$ with quantum numbers $(1,3^*,-1/3)$ and $(1,1,1)$ respectively.
\item $S_3=[(d,u,U);u^c;d^c;U^c]_L$ with quantum numbers $(3,3^*,1/3);\; (3^*,1,-2/3)\; (3^*,1,1/3)$ and $(3^*,1,-2/3)$ respectively.
\item $S_4=[(u,d,D);u^c;d^c;D^c]_L$ with quantum numbers $(3,3,0);\; (3^*,1,-2/3);\; (3^*,1,1/3)$ and $(3^*,1,1/3)$ respectively.
\item $S_5=[(e^-,\nu_e,N_1^0);(E^-,N_2^0,N_3^0);(N_4^0, E^+,e^+)]_L$ with quantum numbers $(1,3^*,-1/3)$; $(1,3^*,-1/3)$ and $(1,3^*,2/3)$ respectively.
\item $S_6=[(\nu_e, e^-,E_1^-);(E^+_2,N_1^0,N_2^0);(N_3^0, E^-_2,E_3^-)$;  $e^+; E_1^+; E_3^+]_L$ 
with quantum numbers $(1,3,-2/3)$; $(1,3,1/3)$; $(1,3,-2/3)$; $(111), (111)$;  and $(111)$ respectively.
\end{itemize}
The different anomalies for these six sets are \cite{Ponce:2002sg} found in Table \ref{tabl1}.
With this table, anomaly-free models, without exotic electric charges can be constructed for one, two or more families. 
As noted in Ref. \cite{Ponce:2002sg}, there are eight three-family models that are anomaly free, which are:
\begin{itemize}
\item Model A: named in the literature \enquote{model  with  right-handed neutrinos}. Its fermion structure is given by
$3S_2+S_3+2S_4$. This model was introduced for first time in the literature  in Ref.~\cite{Singer:1980sw}, rediscovered 
in Refs.~\cite{Montero:1992jk,Foot:1992rh,Foot:1994ym}, with the weak charges presented in Ref.~\cite{Gutierrez:2004sba}.

\item Model B: named in the literature \enquote{ Model with exotic electrons}. This model was 
introduced in the literature in Ref.~\cite{Ozer:1995xi} and its lepton sector was   studied in Ref.~\cite{Ponce:2006au}.
Its fermion structure is given by 
$3S_1+2S_3+S_4$.

\item Model C: named in the literature \enquote{model 
with unique lepton generation one} (three different lepton families). Introduced for the first time  in Ref.~\cite{Ponce:2001jn}
and  its was partially  analyzed in Ref.~\cite{Anderson:2005ab}, where the weak charges only for the  leptons were calculated.
Its fermion structure is given by 
$S_1+S_2+S_3+2S_4+S_5$.
\item Model D:  named in the literature \enquote{model with  unique lepton generation two}. 
Introduced for the first time  in Ref.~\cite{Ponce:2001jn} and  it was partially  analyzed in  Ref.~\cite{Anderson:2005ab},
where the weak charges only for the leptons were calculated. 
Its fermion structure is given by 
$S_1+S_2+2S_3+S_4+S_6$.

\item Model E: we name it as \enquote{model hybrid one} (two different lepton structures).
Its fermion structure is given by 
$S_3+2S_4+2S_5+S_6$.

\item Model F: we name it as \enquote{model hybrid two}.  
Its fermion structure is given by 
$2S_3+S_4+S_5+2S_6$.

\item Model G: we name it as \enquote{ carbon copy one} (three identical families as in the SM).
The fermion structure is the same as the representation of the {\bf 27} of the $E_6$ group \ie $3(S_4+S_5)$. 
The fermion weak charges were presented in the literature in Ref.~\cite{Sanchez:2001ua}
\item Model H: We name it as \enquote{ carbon copy two}. The fermion weak charges for this model were
presented in the literature in Ref.~\cite{Martinez:2001mu}.
Its fermion structure is given by 
$3(S_3+S_6)$.
\end{itemize} 
\begin{center}
\TABLE[ht]{
\noindent\makebox[\textwidth]{
\scalebox{1}{
\label{tabl1}
\begin{tabular}{||l|cccccc||}\hline\hline
Anomalies & $S_1$ & $S_2$ & $S_3$ & $S_4$ & $S_5$ & $S_6$ \\ \hline
$[SU(3)_C]^2U(1)_x$ & 0 & 0 & 0 & 0 & 0 & 0 \\
$[SU(3)_L]^2U(1)_x$ & $-2/3$  & $-1/3$ & 1 & 0& 0 & -1\\
$[Grav]^2U(1)_x$ & 0 & 0 & 0 & 0 & 0 & 0 \\
$[U(1)_x]^3$ & 10/9 & 8/9 & $-12/9$ & $-6/9$& 6/9& 12/9 \\
$[SU(3)_L]^3$ & 1 & $-1$ & $-3$ & 3 & $-3$ & 3\\
\hline\hline
\end{tabular} 
}}
\caption{Anomalies for 3-3-1 fermion fields structures}.
}
\end{center}

\section{Statistical Analysis and Results}
\label{sec:limits}

\begin{center}
\TABLE[ht]{
\noindent\makebox[\textwidth]{
\scalebox{1}{
\label{tab:331limits}
\begin{tabular}{|l|cccc|}
\hline
$Z^{\prime}$& $\mu^{-}\mu^{+}$ &$e^{-}e^{+}$& $l^{-}l^{+}$ & intersection    \\ \hline
\hline
$Z_{331\text{A}}$      &2.36&2.48&2.65& 2.60              \\
$Z_{331\text{B}}$      &2.66&2.72&2.89& 2.88               \\
$Z_{331\text{C}}$      &2.34&2.45&2.57& 2.59               \\  
$Z_{331\text{D}}$      &2.68&2.73&2.91& 2.91               \\ 
$Z_{331\text{E}}$      &2.71&2.71&2.89& 2.87               \\ 
$Z_{331\text{F}}$      &2.67&2.73&2.90& 2.88               \\ 
$Z_{331\text{G}}$      &2.74&2.71&2.92& 2.91               \\
$Z_{331\text{H}}$      &2.65&2.71&2.88& 2.87               \\
$Z_{331\text{minimal}}$&2.68&2.65&2.94& 2.93                   \\
\hline
\end{tabular} }}
\caption{95$\%$ CL lower mass  limits~(in TeV) for some 3-3-1 $Z'$ models.
The second and third columns  contain the 95$\%$ CL lower mass limits  obtained from  the dimuon and dielectron data
in~\cite{Aad:2014cka} respectively~(see the text for details).  In the fourth column appears the 95$\%$ CL lower mass limits  for the combined 
dielectron and dimuon channels. 
Given in the fifth  column  are the lower mass  limits  obtained by finding the intersection of the total cross-section $\sigma^{NLO}$ Eq.~\ref{eq:nlo} 
with the  ATLAS 95$\%$ CL upper limit on the  total cross-section of the $Z_{\text{SSM}}$.}
}
\end{center}
\begin{center}
\TABLE[ht]{
\noindent\makebox[\textwidth]{
\scalebox{1}{
\label{tab:limits}
\begin{tabular}{|l|ccccc|}
\hline
$Z^{\prime}$&  $\mu^{-}\mu^{+}$ &$e^{-}e^{+}$&$l^{-}l^{+}$  &intersection& ATLAS  \\ \hline
\hline
$Z_\chi$~\cite{Robinett:1982tq}                              &2.42&2.48&2.66&2.59 &2.62  \\ 
$Z_\psi$~\cite{Robinett:1982tq}                              &2.20&2.35&2.51&2.42 &2.51  \\
$Z_\eta$~\cite{Witten:1985xc}                                &2.31&2.38&2.56&2.47 &--- \\
$Z_{LR}$~\cite{Pati:1974yy,Mohapatra:1974hk,Mohapatra:1974gc}&2.44&2.54&2.68&2.71 &--- \\
$Z_R$~\cite{Robinett:1982tq}                                 &2.56&2.68&2.87&2.80 &--- \\
$Z_N$~\cite{Ma:1995xk,King:2005jy}                           &2.20&2.36&2.51&2.44 &--- \\
$Z_S$~\cite{Erler:2002pr,Kang:2004pp}                        &2.36&2.42&2.54&2.53 &---\\
$Z_I$~\cite{Robinett:1982tq}                                 &2.31&2.37&2.52&2.48 &--- \\
$Z_{B-L}$~\cite{Pati:1974yy}                                 &2.57&2.68&2.84&2.81 &--- \\
$Z_{\not d}$~\cite{Erler:2011ud}                             &2.75&2.84&2.97&2.94 & --- \\ 
$Z_{\text{SSM}}$                                             &2.57&2.80&2.92&2.91 &2.90  \\
\hline
\end{tabular} }}
\caption{
95$\%$ CL lower mass limits~(in TeV) for  various  $E_6$-motivated $Z'$  models and the SSM.
The second and third columns  contain the 95$\%$ CL lower mass limits  obtained from the dimuon and dielectron data 
in~\cite{Aad:2014cka} respectively~(see the text for details).  In the fourth column appears the 95$\%$ CL lower mass limits  for the combined 
dielectron and dimuon channels. 
Given in the fifth  column  are the lower mass  limits  obtained by finding the intersection of the total cross-section $\sigma^{NLO}$ Eq.~\ref{eq:nlo} 
with the  ATLAS 95$\%$ CL upper limit on the  total cross-section of the $Z_{\text{SSM}}$.
In the sixth  column are the ATLAS published constraints on the respective model.
}}
\end{center}
In reference~\cite{Aad:2014cka} the ATLAS detector at the Large Hadron Collider was used to search 
for high-mass resonances decaying to dielectron or dimuon final 
states.  The experiment analyze proton-proton collisions at a center of mass energy of 8~TeV and
a integrated luminosity of 20.3 $fb^{-1}$ in the dielectron channel 
and  20.5 $fb^{-1}$ in the dimuon channel. From this data they report 95$\%$ CL upper limits on the total cross-section of  
$Z'$ decaying to dilepton final states in $pp$ collisions.
In the aforementioned work  the ATLAS collaboration reported  limits for 
$Z_{\chi}$  and $Z_{\psi}$, which are  $E_6$-motivated $Z'$ models, and for 
the   Sequential Standard Model~(SSM) $Z'_{\text{SSM}}$, which is a model with   couplings to the SM fermions  identical to the   $Z$.
Part of the purpose of this work is to extend 
this analysis to  3-3-1 models  and also to  the remaining $E_6$  models 
which were not considered by ATLAS.  
In this vein we also carry out our own statistical analysis 
by using  a binned likelihood function. The likelihood function is defined as the product of the Poisson probabilities 
over all the dilepton invariant mass bins, \ie
\begin{align}
L(\vec{n}|\vec{\mu}) \equiv \prod\frac{e^{-\mu_i}\mu_i^{n_i}}{n_i!}. 
\end{align}
The confidence levels limits  correspond to   contours of constant  Log-Likelihood Ratio $\text{LLR}(M_{Z'})$, with
  \begin{align}
\label{eq:chimin}
        \text{LLR}(M_{Z'}) =-2\log\frac{L(\vec{n}|\vec{\mu}')}{L(\vec{n}|\vec{\mu})} 
                   = 2\sum_i \left( \mu_i' - \mu_i + n_i\ln \frac{\mu_i}{\mu_i'} \right),
\end{align}
where    $n_i$ is the observed number of events in every bin,
 $\mu_i$ and   $\mu_i'$ are the expected number of events in every bin for the SM and the SM extended by a $Z'$ respectively.
 The explicit expression for the expected number of events is given by
 \begin{align}
\mu_i = K_i\int_{\text{bin}}\frac{d\sigma^{\text{NLO}}}{d\M},
\end{align}
where $K_i$ stand for all the correction factors  necessary to get the expected number of events in every bin.
This corrections  include final state  radiation corrections,  dilepton invariant mass resolution effects,
NNLO QCD, acceptance and efficiency correction factors. We got the $K_i$ from the
ratio between the published SM values for $\mu_i$ from Fig.~2 in~\cite{Aad:2014cka} over  
the  NLO cross-section in the SM,  $\sigma^{\text{NLO}}$ from Eq.~\ref{eq:nlo}, in every bin.
In the calculation of the expected number of events we only took into account the  
couplings of the $Z'$  to the  SM fermions.
In order to find the  95$\%$ CL limits on the masses  for  $E_6$-motivated $Z'$ models  we fix the $Z'$ coupling strength 
to $g_2= 0.4615$ (see Eq.~\ref{eq:hnc} for the $g_2$ definition) 
 and $g_2 = 0.7433$ for 3-3-1 models and the sequential standard model $Z_{\text{SSM}}$.
In our calculation we fix to zero  the mixing angle between the $Z$ and the $Z'$  
in agreement with the most recent  constraints~\cite{Erler:2009jh,Erler:2009ut,delAguila:2010mx,Erler:2010uy}.
It is important to notice that despite the fact that  the number of observed events in every bin is 
Poisson distributed,  according with the Wilks's theorem   the minimum of the likelihood ratio as a function of the 
$Z'$ mass,  follows a $\chi^2$ distribution with degrees of freedom equal to the 
difference of the number of parameters between the two models\footnote{provided  that certain regularity conditions are met.}~\cite{Agashe:2014kda}.
So, the one-parameter 95$\%$ CL  limits  correspond to  LLR$-\text{LLR}_{\text{min}}=3.84$, where 
$\text{LLR}_{\text{min}}$ is the minimum of the LLR as a function of the $Z'$ mass.
For this analysis we used  the thirty five   high-invariant-mass bins  for which the statistical errors are dominant, 
we did not  include low-invariant-mass bins because  other uncertainties become important\footnote{For example
at low-invariant-mass the  theoretical uncertainties become larger than the statistical ones.}.
Following ATLAS, the  bin width is constant in $\log M_{l^{+}l^{-}}$; \ie the border 
between two adjacent  bins, $M_{l^{+}l^{-}}^i$, is given by an exponential function 
$M_{l^{+}l^{-}}^i=M_{l^{+}l^{-}}^1 \exp [(i-1)\times\text{constant}]$,  where $M_{l^{+}l^{-}}^1$ is the leftmost invariant mass value 
and $i =1,2,\cdots$. We fit the ATLAS invariant mass coordinates   to this functional form, getting a good agreement. 

\FIGURE[t]{\label{fig:projected}
\includegraphics[scale=0.6]{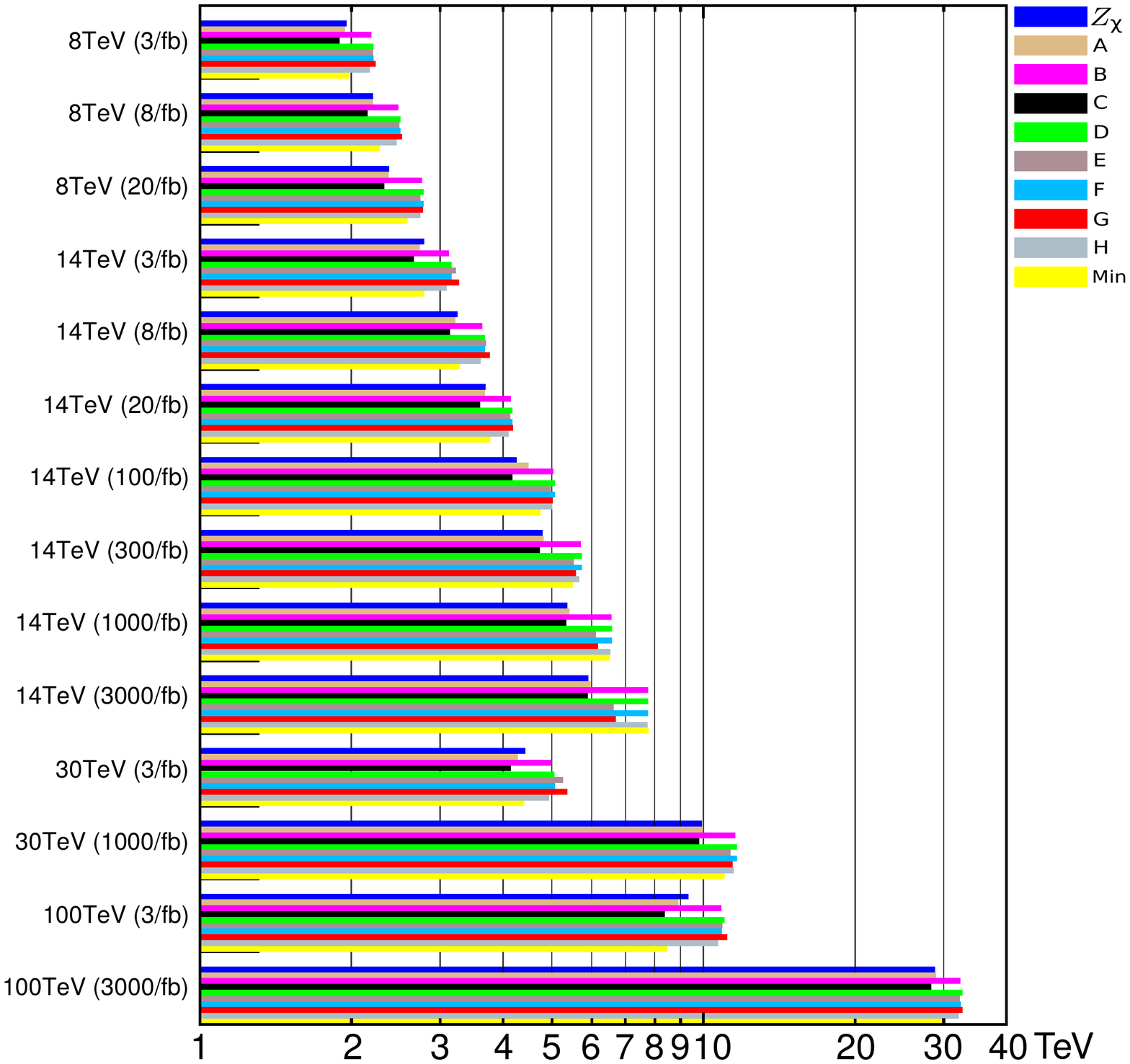}
\caption{Projected 95$\%$ CL exclusion limits  on $M_{Z'}$ for several 3-3-1 models by using our 
statistical methods.  We obtain this limits  by assuming that the number  of observed events $n_i$
is equal to the SM expectation $\mu_i$ in every bin.
We have assumed 
for the product  acceptance$\times$efficiency the ATLAS result for the  dimuon channel as is  
shown in Fig.~1 in Ref.~\cite{Aad:2014cka}.}}
\FIGURE[t]{\label{fig:projected2}
\includegraphics[scale=0.6]{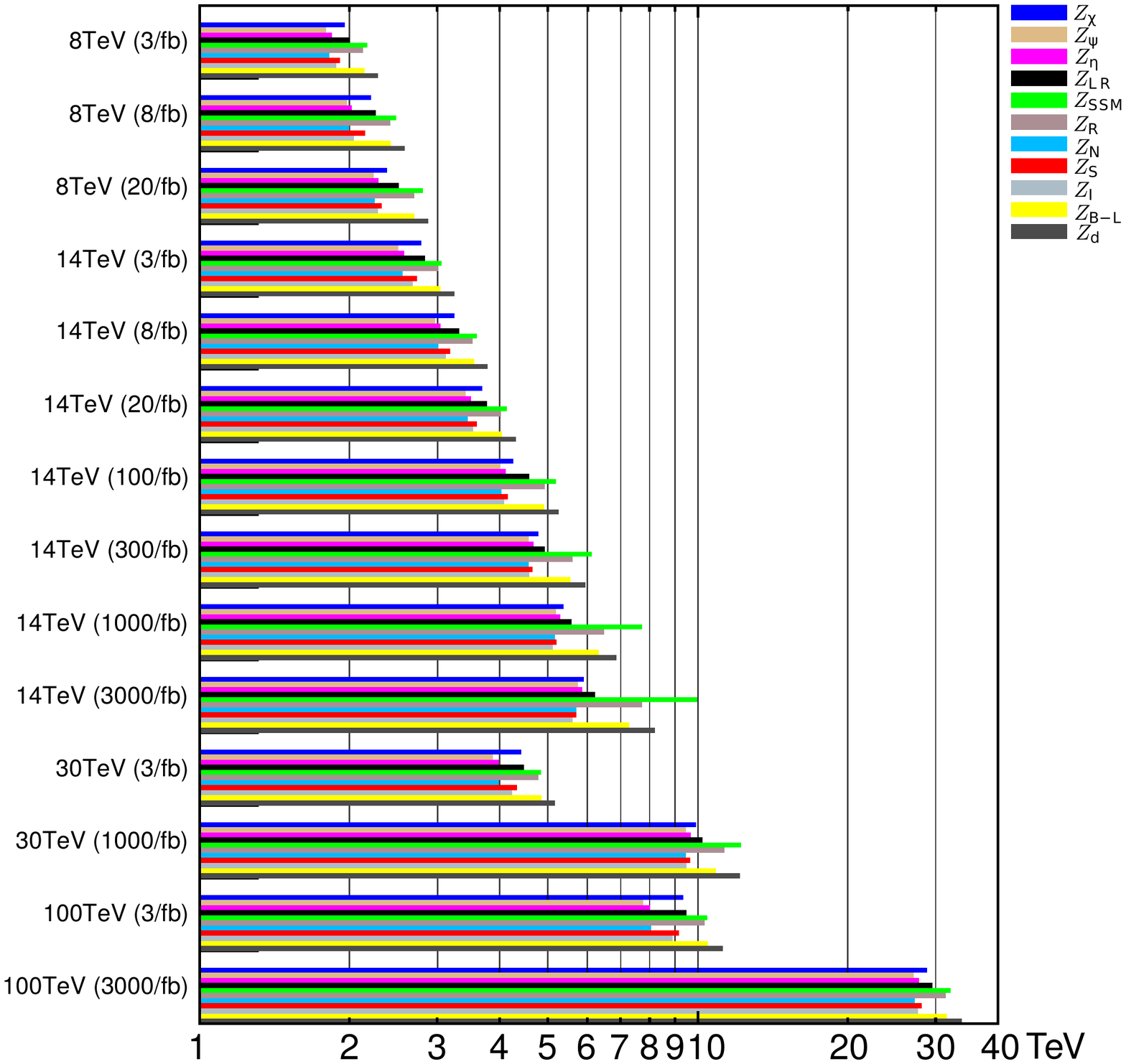}
\caption{Projected 95$\%$ CL exclusion limits  on $M_{Z'}$ for several $E_6$-motivated $Z'$  models and the SSM by using our 
statistical methods.  We obtain this limits  by assuming that the number  of observed events $n_i$
is equal to the SM expectation $\mu_i$ in every bin.
We have assumed 
for the product  acceptance$\times$efficiency the ATLAS result for the  dimuon channel as is  
shown in Fig.~1 in Ref.~\cite{Aad:2014cka}.}}

 The results of this analysis are shown
in Table~\ref{tab:331limits} and Table~\ref{tab:limits}
 where the 95$\%$ CL lower  mass limits for   some  3-3-1 models and various  $E_6$-motivated $Z'$  models
are shown. 	
The second and third columns  contain the 95$\%$ CL lower mass  limits  obtained from the dimuon and dielectron data.
In  our analysis the lower bounds for the $Z'$ mass of the SSM are  2.57~TeV in the dimuon channel and 2.80~TeV in the 
dielectron channel,  which are in good agreement with 
the quoted limits by ATLAS for this model \ie 2.53~TeV in the dimuon channel and   2.79~TeV in the dielectron channel.
 In the fourth column appears the 95$\%$ CL lower mass  limits  for the combined 
 channels. In order to combine the dielectron and dimuon data we add 
 the respective LLR, neglecting systematic uncertainties and  correlations. The validity of 
 this procedure only depends on the validity of the results. As we can see 
 in Table~\ref{tab:limits} for the models $Z_\chi$, $Z_\psi$ and $Z_{\text{SSM}}$
 we obtain 2.66~TeV, 2.51~TeV  and 2.92~TeV which differs at most  1.5$\%$  with  the corresponding  
  ATLAS results 2.62~TeV, 2.51~TeV and 2.90~TeV  respectively.  
  
 In order to make a cross-check of our analysis we make an alternative calculation of the lower bounds.
As  can be seen  from Fig.~5 in Ref.~\cite{Aad:2014cka}, for narrow width resonances the 95$\%$ CL upper limits on the total cross-section of signal events 
is almost model independent for $Z'$ masses below  2~TeV. For larger masses, the constraints are model dependent. 
Since ATLAS does not report upper limits for all the models, 
a useful approximation in the   2-3~TeV range  is to read   the $Z'$  mass lower limit  at the intersection of the theoretical 
total cross-section\footnote{In order to obtain $\sigma^{NLO}$ it is necessary to integrate Eq.~\ref{eq:nlo}}
$\sigma^{NLO}(pp\rightarrow Z' \rightarrow l^{+}l^{-})$ 
Eq.~\ref{eq:nlo}
with the 95$\%$ CL upper limit on the total cross-section of a narrow width resonance. Here, we use 
the upper limit on the total cross-section of  the $Z_{\text{SSM}}$
model in Fig.~5 of~\cite{Aad:2014cka} which was the usual choice in earlier literature~(see for example~\cite{Aad:2012hf}). 
This approximation
allows to  estimate  the 95$\%$ CL lower mass limits differing  from the  corresponding LHC ones  
in at most   a few percent as can be seen in  the fifth column in Table~\ref{tab:limits}.

In Fig.~\ref{fig:projected} the projected 95$\%$ CL exclusion limits  on $M_{Z'}$ for several 3-3-1 models 
are shown.   We obtain this limits  by assuming that the number  of observed events $n_i$
is equal to the SM expectation $\mu_i$ in every bin.  In order to obtain  the bin size
for every center  of mass energy and luminosity in Fig.~\ref{fig:projected},
we took ten  high-invariant-mass  bins  and varied the bin
size until the mass limit reaches a maximum. To obtain the  limits listed there 
we use 30 bins.
We also have assumed 
for the product  acceptance$\times$efficiency the ATLAS result for the  dimuon channel as is  shown in Fig. 1 in \cite{Aad:2014cka}.
The limits in Fig.~\ref{fig:projected2} are comparable with the limits published in~\cite{Godfrey:2013eta}.

\vspace{1cm}
\section{Conclusions}
\label{sec:conclusions}
In the present work we have reported the vector and axial charges for all 
 3-3-1 models  without exotic electric charges for leptons, 
known in the literature. To the best of our knowledge  most of this charges were not reported before
and represent a original contribution to the field. 
By using ATLAS data  from   the Drell-Yang process $pp\rightarrow Z,\gamma \rightarrow l^{+}l^{-} $  
we set 95$\%$ CL lower limits  for the $Z'$  mass  in every one of this models.
We calculated this limits for the dimuon and  the dielectron channels. 
Our results are in accordance with the ATLAS reported   lower  mass limits  for the SSM,  $Z_{\text{SSM}}$ in  every channel. 
By neglecting  systematic uncertainties we were able to combine the two channels  finding 
good agreement with the ATLAS published results. As far as we know 
this is the first time that 3-3-1 models have been constrained with LHC data from ATLAS. 
In addition we also calculated 95$\%$ projected exclusion limits for the 
forthcoming LHC and VLHC  energies and luminosities. As we already 
mentioned in the text, this projected limits are comparable with previous 
calculations, in particular we find that for a center of mass energy of 
14~TeV  and a integrated luminosity of 100fb$^{-1}$ the projected 95$\%$ CL exclusion limits for 3-3-1 models
are between 4~TeV and 5~TeV. 
Part of our long term goal  is to present 
a unified  phenomenological analysis for the 3-3-1 models in oder to set the 
relevance of the forthcoming experiments for every point in the parameter space 
$g_2$ Vs $M_{Z'}$.  We postpone  to a future work a comparative study between FCNC  against 
those coming of direct searches  at hadron colliders.
\section*{Acknowledgments} 
We thank financial support from 
\enquote{Patrimonio Autónomo Fondo Nacional de Financiamiento para la Ciencia, la Tecnología
y la Innovación, Francisco José de Caldas} 
and \enquote{Sostenibilidad-UDEA 2014-2015}.
 R. H. B. thanks to \enquote{Centro de Investigaciones del ITM}.
E.R. acknowledges financial support  provided by
FAPESP. We also like to thank P. Langacker for his valuable suggestions. 

\appendix
\section{Differential Cross-Section}
\label{app:formalism}

The NLO differential cross-section for the DY process with a neutral gauge boson $G$ as the mediator,
$pp\rightarrow GX\rightarrow l^+ l^- X$, 
is given as~\cite{Erler:2011iw,Ellis:1991qj}\footnote{The integration over $z$ is carried out as $\int_0^{(1+\epsilon)} dz \delta(1-z)=1.$},
\bea
\label{eq:nlo}
\frac{d\sigma^{\sbx{0.6}{NLO}}}{d\M\ \ }
&=&\frac{2 }{N_c s}\M\int dzdx_{1}\frac{1}{x_1z}\theta\left(1-\frac{1}{x_{1}zr^{2}_z}
\right)
\sum_{q} \hat{\sigma}_{q\bar{q} \to \ell^+\ell^-}(\M^2)\\\nn
&\times&\Big[\Big\{f_{q}^{A}(x_{1},\M^2)f_{\bar{q}}^{B}(x_{2},\M^2)+f_{\bar{q}}^{A}(x_{1},M^2)f_{q}^{B}(x_{2},\M^2)\Big\}\\\nn
&\times & \Big\{\delta(1-z)+\frac{\alpha_{s}(\M^2)}{2\pi}D_{q}(z)\Big\}\\\nn
&+&\Big\{f_{g}^{A}(x_{1},\M^2)[f_{q}^{B}(x_{2},\M^2)+f_{\bar{q}}^{B}(x_{2},\M^2)]\\\nn
&+&\ f_{g}^{B}(x_{2},M^2)[f_{q}^{A}(x_{1},\M^2)+f_{\bar{q}}^{A}(x_{1},\M^2)]\Big\}
\times \frac{\alpha_{s}(\M^2)}{2\pi}D_{g}(z)\Big],
\eea
where $N_c=3$ is the color factor, $\M$ is the invariant mass of the observed lepton pair and $\sqrt{s}$ is the energy of 
the $p\bar{p}$ collision in the CM frame, $r_z \equiv \sqrt{s}/\M$, and $x_2^{-1} \equiv x_1z r_z ^2$. 
$f_{q/g}^A$ are the PDFs of the quarks and gluons coming from hadron $A$. 
$\alpha_s$ is the strong coupling constant, and
\bea
D_{q}(z)&=&C_{F}\Big[4(1+z^{2})\Big\{\frac{\log(1-z)}{1-z}\Big\}_{+}-2\frac{1+z^{2}}{1-z}\log z
         +\delta(1-z)\Big\{\frac{2\pi^2}{3}-8\Big\}\Big],\\\nn
D_{g}(z)&=&T_{R}\Big[\Big\{z^2+(1-z)^2\Big\}\log\frac{(1-z)^2}{z}+\frac{1}{2}+3z-\frac{7}{2}z^2\Big], 
\eea
with  $C_F=4/3$ and $T_R=1/2$, and the `+' distribution defined as
\be
\int_{0}^{1}dz g(z)\Big\{\frac{\log(1-z)}{1-z}\Big\}_{+}\equiv\int_{0}^{1}dz \Big\{g(z)-g(1)\Big\}\Big\{\frac{\log(1-z)}{1-z}\Big\}.
\ee
At parton level, the expression for the hard scattering cross-section of the process $q\bar{q} \to \ell^+ \ell^-$, is
\be
\label{eq:hard}
\hat{\sigma}_{q\bar{q} \to \ell^+\ell^-} (M^2)= \int_{-1}^{1}\frac{d\hat{\sigma}}{d\cos \theta^*}d\cos \theta^* 
\ee
$$
= \int_{-1}^{1} \frac{d\cos \theta^*}{ 128\pi M^2}\Big\{
\left(\lvert A_{LL}\rvert^2+\lvert A_{RR}\rvert^2\right)(1+\cos\theta^*)^2
+\left(\lvert A_{LR}\rvert^2+\lvert A_{RL}\rvert^2\right)(1-\cos\theta^*)^2\Big\},
$$
where  $\theta^*$ is the polar angle in the CM frame, and 
\be
\label{eq:amplitud}
   A_{ij} = - Q(q) e^2 
             + \frac{g_1^2\, \epsilon_{1i}(q) \epsilon_{1j}(\ell) M^2 }{ M^2 - M_Z^2 + i M_Z \Gamma_Z}
             + \frac{g^2_2\, \epsilon_{2i}(q) \epsilon_{2j}(\ell) M^2 }{ M^2 - M_{Z'}^2 + i M_{Z'} \Gamma_{Z'}},
\ee
where $i,j$ run over $L,R$. 
$Q(q)$ is the electric charge of the quark and $e = g \sin\theta_W$. 
$M_{Z,Z'}$ and $\Gamma_{Z,Z'}$ are the masses and total decay widths of the $Z$ and $Z'$ bosons. 
\be 
   \epsilon_{1L} (f) = T_3 (f) - Q(f)\sin^2\theta_W,   \hspace{24pt}  
   \epsilon_{1R} (f) = - Q(f)\sin^2\theta_W,
\ee
are the effective couplings of the ordinary $Z$ to fermion $f$ entering with coupling strength, $g_1 = g/\cos\theta_W =  0.7433$.
As for the $Z'$ coupling strength, for $E_6$ we employ the (one-loop) unification value~\cite{Robinett:1982tq},
$g_2 = \sqrt{5/3}\sin\theta_W g_1 = 0.4615$; for 3-3-1 models  see appendix~\ref{appendixb}.        
The decay width, $\Gamma_{Z'}$, given in eq.~(\ref{eq:amplitud}), is the sum of the partial decay widths of the $Z'$ boson into all the 
fermions it couples to. The partial decay width into a Dirac fermion pair is written as~\cite{Kang:2004bz}
\begin{equation}
\resizebox{1.0\hsize}{!}{$%
\Gamma_{Z'\rightarrow  f \bar{f}}(\M^2) = \frac{g^{2}_2 M_{Z'} }{ {24\pi}} \sqrt{1 - \frac{4 M_f^2 }{ M_{Z'}^2}}\notag\\
\left[ \left( 1-\frac{M_f^2 }{ M_{Z'}^2} \right)(\epsilon_{2L}^2(f) + \epsilon_{2R}^2(f)) - \frac{6 M_f^2 }{ \M^2}\epsilon_{2L}(f)\epsilon_{2R}(f) \right] \frac{\ \M^2}{M_{Z'}^2}, 
$}
\label{eq:ZffD}
\end{equation}
where $M_f$ is the mass of the final-state fermion.
We add the factor $M^2/M_{Z'}^2$ to get an `$\hat{s}$'-dependent $Z^{\prime}$-width~\cite{Baur:2001ze}.
For the range of $M_{Z'}$ of interest here, $M_f \ll M_{Z'}$ for SM fermions, and the 
above expression becomes independent of the fermion masses. 

\section{The 3-3-1 Couplings}
\label{appendixb}


For the SM extended by a $U(1)^\prime$ extra factor, the neutral current interactions of the fermions are described by the Hamiltonian
\begin{align}\label{eq:hnc}
 H_{NC}=&  \sum_{i=1}^2 g_i Z_{i\mu}^0
 \sum_f\bar{f}\gamma^{\mu}\left(\epsilon_{iL}(f)P_L+\epsilon_{iR}(f)P_R\right)f,
 \end{align}
where $f$ runs over all the SM fermions in the low energy Neutral Current (NC) 
effective Hamiltonian $H_{NC}$, and $P_L=(1-\gamma_5)/2$ and $P_R=(1+\gamma_5)/2$. For 3-3-1 models, 
the ralationship between $g_1$ and $g_2$ is model dependent, but for all the cases we can write
 \begin{align}\label{eq:331hnc}
H_{NC} =&\frac{g}{2 \cos \theta_W}\sum_{i=1}^2  Z_{i\mu}^0
 \sum_f\bar{f}\gamma^{\mu}\left(g_{iV}(f)-g_{iA}(f)\gamma_5\right)f,
\end{align}
\noindent
where the chiral couplings $\epsilon_{iL}(f)$ and $\epsilon_{iR}(f)$ are linear combinations of 
the vector $g_{iV}(f)$ and axial $g_{iA}(f)$ charges given by $\epsilon_{iL}(f)=[g_{iV}(f)+g_{iA}(f)]/2$ 
and $\epsilon_{iR}(f)=[g_{iV}(f)-g_{iA}(f)]/2$.

The physical fields in the former expressions are:
\begin{eqnarray*}
 Z_1^\mu&=& \ \ Z^\mu\cos\theta+Z^{\prime\mu}\sin\theta, \\
 Z_2^\mu&=& -Z^\mu\sin\theta+Z^{\prime\mu}\cos\theta,
\end{eqnarray*}
where $Z^\mu$ and $Z^{\prime\mu}$ are the weak basis states such that $Z^{\mu}$ is identified with the neutral 
gauge boson of the SM. At a first approximation we have taken $\theta=0$.

For the numerical calculations we use the expressions in Tables~\ref{tab:amodel} to \ref{tab:minimal}, where most of the values in the 
Tables are being presented for the first time in the literature. We have also used:
$M_W = 80.401$ GeV, $M_Z = 91.188$ GeV,  $\cos\theta_W = M_W/M_Z$,
$\delta= \sqrt{4\cos^2\theta_W-1}$ and $g_1\equiv g/\cos \theta_W = 0.7433$.

\TABLE[htp]{
\noindent\makebox[\textwidth]{
\label{tab:amodel}
\caption{Model A, $\alpha=1,2,3,$ and $i=1,2$}
\begin{tabular}{c c c}
\hline
\hline 
Field & \vectorc & \axialc \\\hline
$\nu_\alpha$ & $(\frac{1}{2}-\sin^2\theta_W)\frac{1}{\delta}$ &  $(\frac{1}{2}-\sin^2\theta_W)\frac{1}{\delta}$ \\
$e_\alpha$ & $-(-\frac{1}{2}+2\sin^2\theta_W)\frac{1}{\delta}$ &  $\frac{1}{2}\frac{1}{\delta}$ \\
$u_i$ & $(-\frac{1}{2}+\frac{4}{3}\sin^2\theta_W)\frac{1}{\delta}$ & $-\frac{1}{2}\frac{1}{\delta}$ \\
$u_3$ & $(\frac{1}{2}+\frac{1}{3}\sin^2\theta_W)\frac{1}{\delta}$ & $-(-\frac{1}{2}+\sin^2\theta_W)\frac{1}{\delta}$ \\
$d_i$ & $-(\frac{1}{2}-\frac{1}{3}\sin^2\theta_W)\frac{1}{\delta}$ & $-(\frac{1}{2}-\sin^2\theta_W)\frac{1}{\delta}$\\
$d_3$ & $-(-\frac{1}{2}+\frac{2}{3}\sin^2\theta_W)\frac{1}{\delta}$ &  $\frac{1}{2}\frac{1}{\delta}$\\
\hline
\end{tabular}
}
}



\TABLE[htp]{
\noindent\makebox[\textwidth]{
\label{tab:bmodel}
\caption{Model B, $\alpha=1,2,3,$ and $i=1,2$.}
\begin{tabular}{c c c}
\hline
\hline 
Field & \vectorc & \axialc\\\hline
$\nu_\alpha$ & $-\frac{1}{2}\frac{1}{\delta}$ & $-\frac{1}{2}\frac{1}{\delta}$ \\
$e_\alpha$ & $-(\frac{1}{2}+\sin^2\theta_W)\frac{1}{\delta}$ & $-(\frac{1}{2}-\sin^2\theta_W)\frac{1}{\delta}$ \\
$u_i$ & $(\frac{1}{2}+\frac{1}{3}\sin^2\theta_W)\frac{1}{\delta}$ & $-(-\frac{1}{2}+\sin^2\theta_W)\frac{1}{\delta}$ \\
$u_3$ & $(-\frac{1}{2}+\frac{4}{3}\sin^2\theta_W)\frac{1}{\delta}$ & $-\frac{1}{2}\frac{1}{\delta}$ \\
$d_i$ & $(\frac{1}{2}-\frac{2}{3}\sin^2\theta_W)\frac{1}{\delta}$ & $\frac{1}{2}\frac{1}{\delta}$\\
$d_3$ & $(-\frac{1}{2}+\frac{1}{3}\sin^2\theta_W)\frac{1}{\delta}$ & $-(\frac{1}{2}-\sin^2\theta_W)\frac{1}{\delta}$\\
\hline
\end{tabular}}}




\TABLE[htp]{
\noindent\makebox[\textwidth]{
\label{tab:cmodel}
\caption{Model C  and  $i=1,2$}
\centering
\begin{tabular}{c c c}
\hline
\hline 
Field & \vectorc & \axialc\\\hline
$\nu_3$  & $-\frac{1}{2}\frac{1}{\delta}$ &  $-\frac{1}{2}\frac{1}{\delta}$ \\
$\nu_i$ & $-(-\frac{1}{2}+\sin^2\theta_W)\frac{1}{\delta}$ &  $-(-\frac{1}{2}+\sin^2\theta_W)\frac{1}{\delta}$ \\
$e_1$ & $-(\frac{1}{2}+\sin^2\theta_W)\frac{1}{\delta}$ &  $-(\frac{1}{2}-\sin^2\theta_W)\frac{1}{\delta}$ \\
$e_2$ & $-(-\frac{1}{2}+2\sin^2\theta_W)\frac{1}{\delta}$ &  $\frac{1}{2}\frac{1}{\delta}$ \\
$e_3$ & $-3(-\frac{1}{2}+\sin^2\theta_W)\frac{1}{\delta}$ & $(\frac{1}{2}-\cos^2\theta_w)\frac{1}{\delta}$ \\
$u_i$ & $-(\frac{1}{2}-\frac{4}{3}\sin^2\theta_W)\frac{1}{\delta}$ & $-\frac{1}{2}\frac{1}{\delta}$ \\
$u_3$ & $-(-\frac{1}{2}-\frac{1}{3}\sin^2\theta_W)\frac{1}{\delta}$ & $-(-\frac{1}{2}+\sin^2\theta_W)\frac{1}{\delta}$ \\
$d_i$ & $-(\frac{1}{2}-\frac{1}{3}\sin^2\theta_W)\frac{1}{\delta}$ & $-(\frac{1}{2}-\sin^2\theta_W)\frac{1}{\delta}$\\
$d_3$ & $-(-\frac{1}{2}+\frac{2}{3}\sin^2\theta_W)\frac{1}{\delta}$ &  $\frac{1}{2}\frac{1}{\delta}$\\
\hline
\end{tabular}}}



\TABLE[htp]{
\noindent\makebox[\textwidth]{
\label{tab:dmodel}
\caption{Model D   and   $i=1,2$.}
\centering
\begin{tabular}{c c c}
\hline
\hline 
Field & \vectorc & \axialc\\\hline
$\nu_i$  & $-\frac{1}{2}\frac{1}{\delta}$ & $-\frac{1}{2}\frac{1}{\delta}$ \\
$\nu_3$ & $(\frac{1}{2}-\sin^2\theta_W)\frac{1}{\delta}$ &  $(\frac{1}{2}-\sin^2\theta_W)\frac{1}{\delta}$ \\
$e_i$ & $-(\frac{1}{2}+\sin^2\theta_W)\frac{1}{\delta}$ &  $-(\frac{1}{2}-\sin^2\theta_W)\frac{1}{\delta}$ \\
$e_3$ & $-(-\frac{1}{2}+2\sin^2\theta_W)\frac{1}{\delta}$ & $\frac{1}{2}\frac{1}{\delta}$ \\
$u_i$ & $(\frac{1}{2}+\frac{1}{3}\sin^2\theta_W)\frac{1}{\delta}$ & $-(-\frac{1}{2}+\sin^2\theta_W)\frac{1}{\delta}$ \\
$u_3$ & $-(\frac{1}{2}-\frac{4}{3}\sin^2\theta_W)\frac{1}{\delta}$ & $-\frac{1}{2}\frac{1}{\delta}$ \\
$d_i$ & $-(-\frac{1}{2}+\frac{2}{3}\sin^2\theta_W)\frac{1}{\delta}$ & $\frac{1}{2}\frac{1}{\delta}$\\
$d_3$ & $-(\frac{1}{2}-\frac{1}{3}\sin^2\theta_W)\frac{1}{\delta}$ &  $-(\frac{1}{2}-\sin^2\theta_W)\frac{1}{\delta}$\\
\hline
\end{tabular}}}




\TABLE[htp]{
\noindent\makebox[\textwidth]{
\label{tab:emodel}
\caption{Model E and $i=1,2$}
\centering
\begin{tabular}{c c c}
\hline
\hline 
Field & \vectorc & \axialc\\\hline
$\nu_i$  & $(\frac{1}{2}-\sin^2\theta_W)\frac{1}{\delta}$ &  $(\frac{1}{2}-\sin^2\theta_W)\frac{1}{\delta}$ \\
$\nu_3$ & $-\frac{1}{2}\frac{1}{\delta}$ &  $-\frac{1}{2}\frac{1}{\delta}$ \\
$e_i$ & $(\frac{3}{2}-3\sin^2\theta_W)\frac{1}{\delta}$ & $(\frac{1}{2}-\cos^2\theta_W)\frac{1}{\delta}$ \\
$e_3$ & $-(\frac{1}{2}+\sin^2\theta_W)\frac{1}{\delta}$ & $(-\frac{1}{2}+\sin^2\theta_W)\frac{1}{\delta}$ \\
$u_i$ & $(-\frac{1}{2}+\frac{4}{3}\sin^2\theta_W)\frac{1}{\delta}$ & $-\frac{1}{2}\frac{1}{\delta}$ \\
$u_3$ & $(\frac{1}{2}+\frac{1}{3}\sin^2\theta_W)\frac{1}{\delta}$ & $(\frac{1}{2}-\sin^2\theta_W)\frac{1}{\delta}$ \\
$d_i$ & $(-\frac{1}{2}+\frac{1}{3}\sin^2\theta_W)\frac{1}{\delta}$ & $(-\frac{1}{2}+\sin^2\theta_W)\frac{1}{\delta}$\\
$d_3$ & $(\frac{1}{2}-\frac{2}{3}\sin^2\theta_W)\frac{1}{\delta}$ & $\frac{1}{2}\frac{1}{\delta}$ \\
\hline
\end{tabular}}}



\TABLE[htp]{
\noindent\makebox[\textwidth]{
\label{tab:fmodel}
\caption{Model F and   $i=1,2$}
\centering
\begin{tabular}{c c c}
\hline
\hline 
Field & \vectorc & \axialc\\\hline
$\nu_i$  & $-\frac{1}{2}\frac{1}{\delta}$ &  $-\frac{1}{2}\frac{1}{\delta}$ \\
$\nu_3$ &$(\frac{1}{2}-\sin^2\theta_W)\frac{1}{\delta}$ &  $(\frac{1}{2}-\sin^2\theta_W)\frac{1}{\delta}$ \\
$e_i$ & $(-\frac{1}{2}-\sin^2\theta_W)\frac{1}{\delta}$ & $(-\frac{1}{2}+\sin^2\theta_W)\frac{1}{\delta}$ \\
$e_3$ & $(\frac{3}{2}-3\sin^2\theta_W)\frac{1}{\delta}$ & $(\frac{1}{2}-\cos^2\theta_W)\frac{1}{\delta}$ \\
$u_i$ & $(\frac{1}{2}+\frac{1}{3}\sin^2\theta_W)\frac{1}{\delta}$ & $(\frac{1}{2}-\sin^2\theta_W)\frac{1}{\delta}$ \\
$u_3$ & $(-\frac{1}{2}+\frac{4}{3}\sin^2\theta_W)\frac{1}{\delta}$ & $-\frac{1}{2}\frac{1}{\delta}$ \\
$d_i$ & $(\frac{1}{2}-\frac{2}{3}\sin^2\theta_W)\frac{1}{\delta}$ & $\frac{1}{2}\frac{1}{\delta}$\\
$d_3$ & $(-\frac{1}{2}+\frac{1}{3}\sin^2\theta_W)\frac{1}{\delta}$ & $(-\frac{1}{2}+\sin^2\theta_W)\frac{1}{\delta}$  \\
\hline
\end{tabular}}}



\TABLE[htp]{
\noindent\makebox[\textwidth]{
\label{tab:gmodel}
\caption{Model G and $\alpha=1,2,3,$.}
\centering
\begin{tabular}{c c c}
\hline
\hline 
Field & \vectorc & \axialc\\\hline
$\nu_\alpha$ & $(\frac{1}{2}-\sin^2\theta_W)\frac{1}{\delta}$ & $(\frac{1}{2}-\sin^2\theta_W)\frac{1}{\delta}$ \\
$e_\alpha$ &$3(\frac{1}{2}-\sin^2\theta_W)\frac{1}{\delta}$ & $(\frac{1}{2}-\cos^2\theta_W)\frac{1}{\delta}$ \\
$u_\alpha$ & $-(\frac{1}{2}-\frac{4}{3}\sin^2\theta_W)\frac{1}{\delta}$ & $-\frac{1}{2}\frac{1}{\delta}$ \\
$d_\alpha$ & $-(\frac{1}{2}-\frac{1}{3}\sin^2\theta_W)\frac{1}{\delta}$ & $-\frac{1}{2}\cos 2\theta_{W}\frac{1}{\delta}$\\
\hline
\end{tabular}}}


\TABLE[htp]{
\noindent\makebox[\textwidth]{
\label{tab:hmodel}
\caption{Model H  and $\alpha=1,2,3,$. }
\centering
\begin{tabular}{c c c}
\hline
\hline 
Field & \vectorc & \axialc\\\hline
$\nu_\alpha$ & $-\frac{1}{2}\frac{1}{\delta}$ & $-\frac{1}{2}\frac{1}{\delta}$ \\
$e_\alpha$ & $-(\frac{1}{2}+\sin^2\theta_W)\frac{1}{\delta}$ & $(-\frac{1}{2}+\sin^2\theta_W)\frac{1}{\delta}$ \\
$u_\alpha$ & $(\frac{1}{2}+\frac{1}{3}\sin^2\theta_W)\frac{1}{\delta}$ & $(\frac{1}{2}-\sin^2\theta_W)\frac{1}{\delta}$  \\
$d_\alpha$ & $(\frac{1}{2}-\frac{2}{3}\sin^2\theta_W)\frac{1}{\delta}$ & $\frac{1}{2}\frac{1}{\delta}$\\
\hline
\end{tabular}}}
 
\TABLE[htp]{
\noindent\makebox[\textwidth]{
\label{tab:minimal}
\caption{Minimal Model: Pleitez-Frampton~\cite{Dias:2005xj}. $\alpha=1,2,3,$, and $i=1,2$.}
\centering      

\begin{tabular}{c c c}
\hline
\hline 
Field        & \vectorc                                                  & \axialc\\\hline
$\nu_\alpha$ &$-\frac{\sqrt{1-4\sin^2\theta_W}}{2\sqrt{3}}$              & $-\frac{\sqrt{1-4\sin^2\theta_W}}{2\sqrt{3}}$ \\
$e_\alpha$   &$-\frac{\sqrt{3(1-4\sin^2\theta_W)}}{2}$                   & $+\frac{\sqrt{1-4\sin^2\theta_W}}{2\sqrt{3}}$ \\
$u_i$        &$-\frac{-1+6\sin^2\theta_W}{2\sqrt{3(1-4\sin^2\theta_W)}}$ & $+\frac{1+2\sin^2\theta_W}{2\sqrt{3(1-4\sin^2\theta_W)}}$ \\
$t$          &$-\frac{1+4\sin^2\theta_W}{2\sqrt{3(1-4\sin^2\theta_W)}}$  & $-\frac{1-4\sin^2\theta_W}{\sqrt{3(1-4\sin^2\theta_W)}}$ \\ 
$d_i$        &$+\frac{1}{2\sqrt{3(1-4\sin^2\theta_W)}}$                  & $-\frac{-1+4\sin^2\theta_W}{2\sqrt{3(1-4\sin^2\theta_W)}}$\\
$b$          &$-\frac{1-2\sin^2\theta_W}{2\sqrt{3(1-4\sin^2\theta_W)}}$  & $-\frac{1+2\sin^2\theta_W}{2\sqrt{3(1-4\sin^2\theta_W)}}$\\
\hline
\end{tabular}
}} 
\FloatBarrier

\vspace{-10cm}
\bibliographystyle{apsrev4-1}
\bibliography{references331}

\begin{thebibliography}{10}%
\makeatletter
\providecommand \@ifxundefined [1]{%
 \ifx #1\undefined \expandafter \@firstoftwo
 \else \expandafter \@secondoftwo
\fi
}%
\providecommand \@ifnum [1]{%
 \ifnum #1\expandafter \@firstoftwo
 \else \expandafter \@secondoftwo
\fi
}%
\providecommand \enquote [1]{``#1''}%
\providecommand \bibnamefont  [1]{#1}%
\providecommand \bibfnamefont [1]{#1}%
\providecommand \citenamefont [1]{#1}%
\providecommand\href[0]{\@sanitize\@href}%
\providecommand\@href[1]{\endgroup\@@startlink{#1}\endgroup\@@href}%
\providecommand\@@href[1]{#1\@@endlink}%
\providecommand \@sanitize [0]{\begingroup\catcode`\&12\catcode`\#12\relax}%
\@ifxundefined \pdfoutput {\@firstoftwo}{%
 \@ifnum{\z@=\pdfoutput}{\@firstoftwo}{\@secondoftwo}%
}{%
 \providecommand\@@startlink[1]{\leavevmode\special{html:<a href="#1">}}%
 \providecommand\@@endlink[0]{\special{html:</a>}}%
}{%
 \providecommand\@@startlink[1]{%
  \leavevmode
  \pdfstartlink
   attr{/Border[0 0 1 ]/H/I/C[0 1 1]}%
   user{/Subtype/Link/A<</Type/Action/S/URI/URI(#1)>>}%
  \relax
 }%
 \providecommand\@@endlink[0]{\pdfendlink}%
}%
\providecommand \url  [0]{\begingroup\@sanitize \@url }%
\providecommand \@url [1]{\endgroup\@href {#1}{\urlprefix}}%
\providecommand \urlprefix [0]{URL }%
\providecommand \Eprint[0]{\href }%
\@ifxundefined \urlstyle {%
  \providecommand \doi [1]{doi:\discretionary{}{}{}#1}%
}{%
  \providecommand \doi [0]{doi:\discretionary{}{}{}\begingroup
  \urlstyle{rm}\Url }%
}%
\providecommand \doibase [0]{http://dx.doi.org/}%
\providecommand \Doi[1]{\href{\doibase#1}}%
\providecommand \bibAnnote [3]{%
  \BibitemShut{#1}%
  \begin{quotation}\noindent
    \textsc{Key:}\ #2\\\textsc{Annotation:}\ #3%
  \end{quotation}%
}%
\providecommand \bibAnnoteFile [2]{%
  \IfFileExists{#2}{\bibAnnote {#1} {#2} {\input{#2}}}{}%
}%
\providecommand \typeout [0]{\immediate \write \m@ne }%
\providecommand \selectlanguage [0]{\@gobble}%
\providecommand \bibinfo [0]{\@secondoftwo}%
\providecommand \bibfield [0]{\@secondoftwo}%
\providecommand \translation [1]{[#1]}%
\providecommand \BibitemOpen[0]{}%
\providecommand \bibitemStop [0]{}%
\providecommand \bibitemNoStop [0]{.\EOS\space}%
\providecommand \EOS [0]{\spacefactor3000\relax}%
\providecommand \BibitemShut [1]{\csname bibitem#1\endcsname}%
\bibitem{Donoghue:1992dd}%
  \BibitemOpen
  \bibfield{author}{%
  \bibinfo {author} {\bibfnamefont{J.}~\bibnamefont{Donoghue}}, \bibinfo
  {author} {\bibfnamefont{E.}~\bibnamefont{Golowich}},\ and\ \bibinfo {author}
  {\bibfnamefont{B.~R.}\ \bibnamefont{Holstein}},\ }%
  \emph{\bibinfo {title} {{Dynamics of the standard model}}}\ (\bibinfo
  {publisher} {Cambridge University Press, Cambridge, U.K},\ \bibinfo {year}
  {1992})%
  \bibAnnoteFile{NoStop}{Donoghue:1992dd}%
\bibitem{Langacker:2008yv}%
  \BibitemOpen
  \bibfield{author}{%
  \bibinfo {author} {\bibfnamefont{P.}~\bibnamefont{Langacker}},\ }%
  \bibfield{journal}{%
  \Doi{10.1103/RevModPhys.81.1199}{\bibinfo {journal} {Rev.Mod.Phys.}}\ }%
  \textbf{\bibinfo {volume} {81}},\ \bibinfo {pages} {1199} (\bibinfo {year}
  {2009}),\ \Eprint{http://arxiv.org/abs/0801.1345}{arXiv:0801.1345 [hep-ph]}%
  \bibAnnoteFile{NoStop}{Langacker:2008yv}%
\bibitem{Slansky:1981yr}%
  \BibitemOpen
  \bibfield{author}{%
  \bibinfo {author} {\bibfnamefont{R.}~\bibnamefont{Slansky}},\ }%
  \bibfield{journal}{%
  \Doi{10.1016/0370-1573(81)90092-2}{\bibinfo {journal} {Phys.Rept.}}\ }%
  \textbf{\bibinfo {volume} {79}},\ \bibinfo {pages} {1} (\bibinfo {year}
  {1981})%
  \bibAnnoteFile{NoStop}{Slansky:1981yr}%
\bibitem{Langacker:2000ju}%
  \BibitemOpen
  \bibfield{author}{%
  \bibinfo {author} {\bibfnamefont{P.}~\bibnamefont{Langacker}}\ and\ \bibinfo
  {author} {\bibfnamefont{M.}~\bibnamefont{Plumacher}},\ }%
  \bibfield{journal}{%
  \Doi{10.1103/PhysRevD.62.013006}{\bibinfo {journal} {Phys. Rev.}}\ }%
  \textbf{\bibinfo {volume} {D62}},\ \bibinfo {pages} {013006} (\bibinfo {year}
  {2000}),\ \Eprint{http://arxiv.org/abs/hep-ph/0001204}{arXiv:hep-ph/0001204
  [hep-ph]}%
  \bibAnnoteFile{NoStop}{Langacker:2000ju}%
\bibitem{Appelquist:2002mw}%
  \BibitemOpen
  \bibfield{author}{%
  \bibinfo {author} {\bibfnamefont{T.}~\bibnamefont{Appelquist}}, \bibinfo
  {author} {\bibfnamefont{B.~A.}\ \bibnamefont{Dobrescu}},\ and\ \bibinfo
  {author} {\bibfnamefont{A.~R.}\ \bibnamefont{Hopper}},\ }%
  \bibfield{journal}{%
  \Doi{10.1103/PhysRevD.68.035012}{\bibinfo {journal} {Phys.Rev.}}\ }%
  \textbf{\bibinfo {volume} {D68}},\ \bibinfo {pages} {035012} (\bibinfo {year}
  {2003}),\ \Eprint{http://arxiv.org/abs/hep-ph/0212073}{arXiv:hep-ph/0212073
  [hep-ph]}%
  \bibAnnoteFile{NoStop}{Appelquist:2002mw}%
\bibitem{Carena:2004xs}%
  \BibitemOpen
  \bibfield{author}{%
  \bibinfo {author} {\bibfnamefont{M.~S.}\ \bibnamefont{Carena}}, \bibinfo
  {author} {\bibfnamefont{A.}~\bibnamefont{Daleo}}, \bibinfo {author}
  {\bibfnamefont{B.~A.}\ \bibnamefont{Dobrescu}},\ and\ \bibinfo {author}
  {\bibfnamefont{T.~M.}\ \bibnamefont{Tait}},\ }%
  \bibfield{journal}{%
  \Doi{10.1103/PhysRevD.70.093009}{\bibinfo {journal} {Phys.Rev.}}\ }%
  \textbf{\bibinfo {volume} {D70}},\ \bibinfo {pages} {093009} (\bibinfo {year}
  {2004}),\ \Eprint{http://arxiv.org/abs/hep-ph/0408098}{arXiv:hep-ph/0408098
  [hep-ph]}%
  \bibAnnoteFile{NoStop}{Carena:2004xs}%
\bibitem{Gursey:1975ki}%
  \BibitemOpen
  \bibfield{author}{%
  \bibinfo {author} {\bibfnamefont{F.}~\bibnamefont{Gursey}}, \bibinfo {author}
  {\bibfnamefont{P.}~\bibnamefont{Ramond}},\ and\ \bibinfo {author}
  {\bibfnamefont{P.}~\bibnamefont{Sikivie}},\ }%
  \bibfield{journal}{%
  \Doi{10.1016/0370-2693(76)90417-2}{\bibinfo {journal} {Phys.Lett.}}\ }%
  \textbf{\bibinfo {volume} {B60}},\ \bibinfo {pages} {177} (\bibinfo {year}
  {1976})%
  \bibAnnoteFile{NoStop}{Gursey:1975ki}%
\bibitem{Robinett:1982tq}%
  \BibitemOpen
  \bibfield{author}{%
  \bibinfo {author} {\bibfnamefont{R.}~\bibnamefont{Robinett}}\ and\ \bibinfo
  {author} {\bibfnamefont{J.~L.}\ \bibnamefont{Rosner}},\ }%
  \bibfield{journal}{%
  \Doi{10.1103/PhysRevD.26.2396}{\bibinfo {journal} {Phys.Rev.}}\ }%
  \textbf{\bibinfo {volume} {D26}},\ \bibinfo {pages} {2396} (\bibinfo {year}
  {1982})%
  \bibAnnoteFile{NoStop}{Robinett:1982tq}%
\bibitem{Erler:2011ud}%
  \BibitemOpen
  \bibfield{author}{%
  \bibinfo {author} {\bibfnamefont{J.}~\bibnamefont{Erler}}, \bibinfo {author}
  {\bibfnamefont{P.}~\bibnamefont{Langacker}}, \bibinfo {author}
  {\bibfnamefont{S.}~\bibnamefont{Munir}},\ and\ \bibinfo {author}
  {\bibfnamefont{E.}~\bibnamefont{Rojas}},\ }%
  \bibfield{journal}{%
  \Doi{10.1007/JHEP11(2011)076}{\bibinfo {journal} {JHEP}}\ }%
  \textbf{\bibinfo {volume} {1111}},\ \bibinfo {pages} {076} (\bibinfo {year}
  {2011}),\ \Eprint{http://arxiv.org/abs/1103.2659}{arXiv:1103.2659 [hep-ph]}%
  \bibAnnoteFile{NoStop}{Erler:2011ud}%
\bibitem{Singer:1980sw}%
  \BibitemOpen
  \bibfield{author}{%
  \bibinfo {author} {\bibfnamefont{M.}~\bibnamefont{Singer}}, \bibinfo {author}
  {\bibfnamefont{J.}~\bibnamefont{Valle}},\ and\ \bibinfo {author}
  {\bibfnamefont{J.}~\bibnamefont{Schechter}},\ }%
  \bibfield{journal}{%
  \Doi{10.1103/PhysRevD.22.738}{\bibinfo {journal} {Phys.Rev.}}\ }%
  \textbf{\bibinfo {volume} {D22}},\ \bibinfo {pages} {738} (\bibinfo {year}
  {1980})%
  \bibAnnoteFile{NoStop}{Singer:1980sw}%
\bibitem{Pisano:1991ee}%
  \BibitemOpen
  \bibfield{author}{%
  \bibinfo {author} {\bibfnamefont{F.}~\bibnamefont{Pisano}}\ and\ \bibinfo
  {author} {\bibfnamefont{V.}~\bibnamefont{Pleitez}},\ }%
  \bibfield{journal}{%
  \Doi{10.1103/PhysRevD.46.410}{\bibinfo {journal} {Phys.Rev.}}\ }%
  \textbf{\bibinfo {volume} {D46}},\ \bibinfo {pages} {410} (\bibinfo {year}
  {1992}),\ \Eprint{http://arxiv.org/abs/hep-ph/9206242}{arXiv:hep-ph/9206242
  [hep-ph]}%
  \bibAnnoteFile{NoStop}{Pisano:1991ee}%
\bibitem{Frampton:1992wt}%
  \BibitemOpen
  \bibfield{author}{%
  \bibinfo {author} {\bibfnamefont{P.}~\bibnamefont{Frampton}},\ }%
  \bibfield{journal}{%
  \Doi{10.1103/PhysRevLett.69.2889}{\bibinfo {journal} {Phys.Rev.Lett.}}\ }%
  \textbf{\bibinfo {volume} {69}},\ \bibinfo {pages} {2889} (\bibinfo {year}
  {1992})%
  \bibAnnoteFile{NoStop}{Frampton:1992wt}%
\bibitem{Montero:1992jk}%
  \BibitemOpen
  \bibfield{author}{%
  \bibinfo {author} {\bibfnamefont{J.}~\bibnamefont{Montero}}, \bibinfo
  {author} {\bibfnamefont{F.}~\bibnamefont{Pisano}},\ and\ \bibinfo {author}
  {\bibfnamefont{V.}~\bibnamefont{Pleitez}},\ }%
  \bibfield{journal}{%
  \Doi{10.1103/PhysRevD.47.2918}{\bibinfo {journal} {Phys.Rev.}}\ }%
  \textbf{\bibinfo {volume} {D47}},\ \bibinfo {pages} {2918} (\bibinfo {year}
  {1993}),\ \Eprint{http://arxiv.org/abs/hep-ph/9212271}{arXiv:hep-ph/9212271
  [hep-ph]}%
  \bibAnnoteFile{NoStop}{Montero:1992jk}%
\bibitem{Foot:1992rh}%
  \BibitemOpen
  \bibfield{author}{%
  \bibinfo {author} {\bibfnamefont{R.}~\bibnamefont{Foot}}, \bibinfo {author}
  {\bibfnamefont{O.~F.}\ \bibnamefont{Hernandez}}, \bibinfo {author}
  {\bibfnamefont{F.}~\bibnamefont{Pisano}},\ and\ \bibinfo {author}
  {\bibfnamefont{V.}~\bibnamefont{Pleitez}},\ }%
  \bibfield{journal}{%
  \Doi{10.1103/PhysRevD.47.4158}{\bibinfo {journal} {Phys.Rev.}}\ }%
  \textbf{\bibinfo {volume} {D47}},\ \bibinfo {pages} {4158} (\bibinfo {year}
  {1993}),\ \Eprint{http://arxiv.org/abs/hep-ph/9207264}{arXiv:hep-ph/9207264
  [hep-ph]}%
  \bibAnnoteFile{NoStop}{Foot:1992rh}%
\bibitem{Foot:1994ym}%
  \BibitemOpen
  \bibfield{author}{%
  \bibinfo {author} {\bibfnamefont{R.}~\bibnamefont{Foot}}, \bibinfo {author}
  {\bibfnamefont{H.~N.}\ \bibnamefont{Long}},\ and\ \bibinfo {author}
  {\bibfnamefont{T.~A.}\ \bibnamefont{Tran}},\ }%
  \bibfield{journal}{%
  \Doi{10.1103/PhysRevD.50.R34}{\bibinfo {journal} {Phys. Rev.}}\ }%
  \textbf{\bibinfo {volume} {D50}},\ \bibinfo {pages} {34} (\bibinfo {year}
  {1994}),\ \Eprint{http://arxiv.org/abs/hep-ph/9402243}{arXiv:hep-ph/9402243
  [hep-ph]}%
  \bibAnnoteFile{NoStop}{Foot:1994ym}%
\bibitem{Ozer:1995xi}%
  \BibitemOpen
  \bibfield{author}{%
  \bibinfo {author} {\bibfnamefont{M.}~\bibnamefont{Ozer}},\ }%
  \bibfield{journal}{%
  \Doi{10.1103/PhysRevD.54.1143}{\bibinfo {journal} {Phys.Rev.}}\ }%
  \textbf{\bibinfo {volume} {D54}},\ \bibinfo {pages} {1143} (\bibinfo {year}
  {1996})%
  \bibAnnoteFile{NoStop}{Ozer:1995xi}%
\bibitem{Ponce:2001jn}%
  \BibitemOpen
  \bibfield{author}{%
  \bibinfo {author} {\bibfnamefont{W.~A.}\ \bibnamefont{Ponce}}, \bibinfo
  {author} {\bibfnamefont{J.~B.}\ \bibnamefont{Florez}},\ and\ \bibinfo
  {author} {\bibfnamefont{L.~A.}\ \bibnamefont{Sanchez}},\ }%
  \bibfield{journal}{%
  \Doi{10.1142/S0217751X02005815}{\bibinfo {journal} {Int.J.Mod.Phys.}}\ }%
  \textbf{\bibinfo {volume} {A17}},\ \bibinfo {pages} {643} (\bibinfo {year}
  {2002}),\ \Eprint{http://arxiv.org/abs/hep-ph/0103100}{arXiv:hep-ph/0103100
  [hep-ph]}%
  \bibAnnoteFile{NoStop}{Ponce:2001jn}%
\bibitem{Ponce:2002sg}%
  \BibitemOpen
  \bibfield{author}{%
  \bibinfo {author} {\bibfnamefont{W.~A.}\ \bibnamefont{Ponce}}, \bibinfo
  {author} {\bibfnamefont{Y.}~\bibnamefont{Giraldo}},\ and\ \bibinfo {author}
  {\bibfnamefont{L.~A.}\ \bibnamefont{Sanchez}},\ }%
  \bibfield{journal}{%
  \Doi{10.1103/PhysRevD.67.075001}{\bibinfo {journal} {Phys.Rev.}}\ }%
  \textbf{\bibinfo {volume} {D67}},\ \bibinfo {pages} {075001} (\bibinfo {year}
  {2003}),\ \Eprint{http://arxiv.org/abs/hep-ph/0210026}{arXiv:hep-ph/0210026
  [hep-ph]}%
  \bibAnnoteFile{NoStop}{Ponce:2002sg}%
\bibitem{Ozer:1996jc}%
  \BibitemOpen
  \bibfield{author}{%
  \bibinfo {author} {\bibfnamefont{M.}~\bibnamefont{Ozer}},\ }%
  \bibfield{journal}{%
  \Doi{10.1103/PhysRevD.54.4561}{\bibinfo {journal} {Phys.Rev.}}\ }%
  \textbf{\bibinfo {volume} {D54}},\ \bibinfo {pages} {4561} (\bibinfo {year}
  {1996})%
  \bibAnnoteFile{NoStop}{Ozer:1996jc}%
\bibitem{Montero:2005yb}%
  \BibitemOpen
  \bibfield{author}{%
  \bibinfo {author} {\bibfnamefont{J.}~\bibnamefont{Montero}}, \bibinfo
  {author} {\bibfnamefont{C.}~\bibnamefont{Nishi}}, \bibinfo {author}
  {\bibfnamefont{V.}~\bibnamefont{Pleitez}}, \bibinfo {author}
  {\bibfnamefont{O.}~\bibnamefont{Ravinez}},\ and\ \bibinfo {author}
  {\bibfnamefont{M.}~\bibnamefont{Rodriguez}},\ }%
  \bibfield{journal}{%
  \Doi{10.1103/PhysRevD.73.016003}{\bibinfo {journal} {Phys.Rev.}}\ }%
  \textbf{\bibinfo {volume} {D73}},\ \bibinfo {pages} {016003} (\bibinfo {year}
  {2006}),\ \Eprint{http://arxiv.org/abs/hep-ph/0511100}{arXiv:hep-ph/0511100
  [hep-ph]}%
  \bibAnnoteFile{NoStop}{Montero:2005yb}%
\bibitem{Buras:2004uu}%
  \BibitemOpen
  \bibfield{author}{%
  \bibinfo {author} {\bibfnamefont{A.~J.}\ \bibnamefont{Buras}}, \bibinfo
  {author} {\bibfnamefont{F.}~\bibnamefont{Schwab}},\ and\ \bibinfo {author}
  {\bibfnamefont{S.}~\bibnamefont{Uhlig}},\ }%
  \bibfield{journal}{%
  \Doi{10.1103/RevModPhys.80.965}{\bibinfo {journal} {Rev.Mod.Phys.}}\ }%
  \textbf{\bibinfo {volume} {80}},\ \bibinfo {pages} {965} (\bibinfo {year}
  {2008}),\ \Eprint{http://arxiv.org/abs/hep-ph/0405132}{arXiv:hep-ph/0405132
  [hep-ph]}%
  \bibAnnoteFile{NoStop}{Buras:2004uu}%
\bibitem{Buras:2012dp}%
  \BibitemOpen
  \bibfield{author}{%
  \bibinfo {author} {\bibfnamefont{A.~J.}\ \bibnamefont{Buras}}, \bibinfo
  {author} {\bibfnamefont{F.}~\bibnamefont{De~Fazio}}, \bibinfo {author}
  {\bibfnamefont{J.}~\bibnamefont{Girrbach}},\ and\ \bibinfo {author}
  {\bibfnamefont{M.~V.}\ \bibnamefont{Carlucci}},\ }%
  \bibfield{journal}{%
  \Doi{10.1007/JHEP02(2013)023}{\bibinfo {journal} {JHEP}}\ }%
  \textbf{\bibinfo {volume} {1302}},\ \bibinfo {pages} {023} (\bibinfo {year}
  {2013}),\ \Eprint{http://arxiv.org/abs/1211.1237}{arXiv:1211.1237 [hep-ph]}%
  \bibAnnoteFile{NoStop}{Buras:2012dp}%
\bibitem{Buras:2013ooa}%
  \BibitemOpen
  \bibfield{author}{%
  \bibinfo {author} {\bibfnamefont{A.~J.}\ \bibnamefont{Buras}}\ and\ \bibinfo
  {author} {\bibfnamefont{J.}~\bibnamefont{Girrbach}},\ }%
  \bibfield{journal}{%
  \Doi{10.1088/0034-4885/77/8/086201}{\bibinfo {journal} {Rept.Prog.Phys.}}\ }%
  \textbf{\bibinfo {volume} {77}},\ \bibinfo {pages} {086201} (\bibinfo {year}
  {2014}),\ \Eprint{http://arxiv.org/abs/1306.3775}{arXiv:1306.3775 [hep-ph]}%
  \bibAnnoteFile{NoStop}{Buras:2013ooa}%
\bibitem{Buras:2014yna}%
  \BibitemOpen
  \bibfield{author}{%
  \bibinfo {author} {\bibfnamefont{A.~J.}\ \bibnamefont{Buras}}, \bibinfo
  {author} {\bibfnamefont{F.}~\bibnamefont{De~Fazio}},\ and\ \bibinfo {author}
  {\bibfnamefont{J.}~\bibnamefont{Girrbach-Noe}},\ }%
  \bibfield{journal}{%
  \Doi{10.1007/JHEP08(2014)039}{\bibinfo {journal} {JHEP}}\ }%
  \textbf{\bibinfo {volume} {1408}},\ \bibinfo {pages} {039} (\bibinfo {year}
  {2014}),\ \Eprint{http://arxiv.org/abs/1405.3850}{arXiv:1405.3850 [hep-ph]}%
  \bibAnnoteFile{NoStop}{Buras:2014yna}%
\bibitem{Promberger:2007py}%
  \BibitemOpen
  \bibfield{author}{%
  \bibinfo {author} {\bibfnamefont{C.}~\bibnamefont{Promberger}}, \bibinfo
  {author} {\bibfnamefont{S.}~\bibnamefont{Schatt}},\ and\ \bibinfo {author}
  {\bibfnamefont{F.}~\bibnamefont{Schwab}},\ }%
  \bibfield{journal}{%
  \Doi{10.1103/PhysRevD.75.115007}{\bibinfo {journal} {Phys.Rev.}}\ }%
  \textbf{\bibinfo {volume} {D75}},\ \bibinfo {pages} {115007} (\bibinfo {year}
  {2007}),\ \Eprint{http://arxiv.org/abs/hep-ph/0702169}{arXiv:hep-ph/0702169
  [HEP-PH]}%
  \bibAnnoteFile{NoStop}{Promberger:2007py}%
\bibitem{Cabarcas:2011hb}%
  \BibitemOpen
  \bibfield{author}{%
  \bibinfo {author} {\bibfnamefont{J.}~\bibnamefont{Cabarcas}}, \bibinfo
  {author} {\bibfnamefont{J.}~\bibnamefont{Duarte}},\ and\ \bibinfo {author}
  {\bibfnamefont{J.-A.}\ \bibnamefont{Rodriguez}},\ }%
  \bibfield{journal}{%
  \Doi{10.1155/2012/657582}{\bibinfo {journal} {Adv.High Energy Phys.}}\ }%
  \textbf{\bibinfo {volume} {2012}},\ \bibinfo {pages} {657582} (\bibinfo
  {year} {2012}),\ \Eprint{http://arxiv.org/abs/1111.0315}{arXiv:1111.0315
  [hep-ph]}%
  \bibAnnoteFile{NoStop}{Cabarcas:2011hb}%
\bibitem{Machado:2013jca}%
  \BibitemOpen
  \bibfield{author}{%
  \bibinfo {author} {\bibfnamefont{A.}~\bibnamefont{Machado}}, \bibinfo
  {author} {\bibfnamefont{J.}~\bibnamefont{Montero}},\ and\ \bibinfo {author}
  {\bibfnamefont{V.}~\bibnamefont{Pleitez}},\ }%
  \bibfield{journal}{%
  \Doi{10.1103/PhysRevD.88.113002}{\bibinfo {journal} {Phys.Rev.}}\ }%
  \textbf{\bibinfo {volume} {D88}},\ \bibinfo {pages} {113002} (\bibinfo {year}
  {2013}),\ \Eprint{http://arxiv.org/abs/1305.1921}{arXiv:1305.1921 [hep-ph]}%
  \bibAnnoteFile{NoStop}{Machado:2013jca}%
\bibitem{Martinez:2014lta}%
  \BibitemOpen
  \bibfield{author}{%
  \bibinfo {author} {\bibfnamefont{R.}~\bibnamefont{Martinez}}\ and\ \bibinfo
  {author} {\bibfnamefont{F.}~\bibnamefont{Ochoa}},\ }%
  \bibfield{journal}{%
  \Doi{10.1103/PhysRevD.90.015028}{\bibinfo {journal} {Phys.Rev.}}\ }%
  \textbf{\bibinfo {volume} {D90}},\ \bibinfo {pages} {015028} (\bibinfo {year}
  {2014}),\ \Eprint{http://arxiv.org/abs/1405.4566}{arXiv:1405.4566 [hep-ph]}%
  \bibAnnoteFile{NoStop}{Martinez:2014lta}%
\bibitem{deSousaPires:1998jc}%
  \BibitemOpen
  \bibfield{author}{%
  \bibinfo {author} {\bibfnamefont{C.~A.}\ \bibnamefont{de~Sousa~Pires}}\ and\
  \bibinfo {author} {\bibfnamefont{O.}~\bibnamefont{Ravinez}},\ }%
  \bibfield{journal}{%
  \Doi{10.1103/PhysRevD.58.035008}{\bibinfo {journal} {Phys.Rev.}}\ }%
  \textbf{\bibinfo {volume} {D58}},\ \bibinfo {pages} {035008} (\bibinfo {year}
  {1998}),\ \Eprint{http://arxiv.org/abs/hep-ph/9803409}{arXiv:hep-ph/9803409
  [hep-ph]}%
  \bibAnnoteFile{NoStop}{deSousaPires:1998jc}%
\bibitem{VanDong:2005ux}%
  \BibitemOpen
  \bibfield{author}{%
  \bibinfo {author} {\bibfnamefont{P.~V.}\ \bibnamefont{Dong}}\ and\ \bibinfo
  {author} {\bibfnamefont{H.~N.}\ \bibnamefont{Long}},\ }%
  \bibfield{journal}{%
  \Doi{10.1142/S0217751X06035191}{\bibinfo {journal} {Int.J.Mod.Phys.}}\ }%
  \textbf{\bibinfo {volume} {A21}},\ \bibinfo {pages} {6677} (\bibinfo {year}
  {2006}),\ \Eprint{http://arxiv.org/abs/hep-ph/0507155}{arXiv:hep-ph/0507155
  [hep-ph]}%
  \bibAnnoteFile{NoStop}{VanDong:2005ux}%
\bibitem{Pal:1994ba}%
  \BibitemOpen
  \bibfield{author}{%
  \bibinfo {author} {\bibfnamefont{P.~B.}\ \bibnamefont{Pal}},\ }%
  \bibfield{journal}{%
  \Doi{10.1103/PhysRevD.52.1659}{\bibinfo {journal} {Phys.Rev.}}\ }%
  \textbf{\bibinfo {volume} {D52}},\ \bibinfo {pages} {1659} (\bibinfo {year}
  {1995}),\ \Eprint{http://arxiv.org/abs/hep-ph/9411406}{arXiv:hep-ph/9411406
  [hep-ph]}%
  \bibAnnoteFile{NoStop}{Pal:1994ba}%
\bibitem{Aaltonen:2007ps}%
  \BibitemOpen
  \bibfield{author}{%
  \bibinfo {author} {\bibfnamefont{T.}~\bibnamefont{Aaltonen}} \emph{et~al.}
  (\bibinfo {collaboration} {CDF Collaboration}),\ }%
  \bibfield{journal}{%
  \Doi{10.1103/PhysRevD.77.112001}{\bibinfo {journal} {Phys.Rev.}}\ }%
  \textbf{\bibinfo {volume} {D77}},\ \bibinfo {pages} {112001} (\bibinfo {year}
  {2008}),\ \Eprint{http://arxiv.org/abs/0708.3642}{arXiv:0708.3642 [hep-ex]}%
  \bibAnnoteFile{NoStop}{Aaltonen:2007ps}%
\bibitem{Erler:2011iw}%
  \BibitemOpen
  \bibfield{author}{%
  \bibinfo {author} {\bibfnamefont{J.}~\bibnamefont{Erler}}, \bibinfo {author}
  {\bibfnamefont{P.}~\bibnamefont{Langacker}}, \bibinfo {author}
  {\bibfnamefont{S.}~\bibnamefont{Munir}},\ and\ \bibinfo {author}
  {\bibfnamefont{E.}~\bibnamefont{Rojas}}}%
   (\bibinfo {year} {2011}),\
  \Eprint{http://arxiv.org/abs/1108.0685}{arXiv:1108.0685 [hep-ph]}%
  \bibAnnoteFile{NoStop}{Erler:2011iw}%
\bibitem{Lai:2010vv}%
  \BibitemOpen
  \bibfield{author}{%
  \bibinfo {author} {\bibfnamefont{H.-L.}\ \bibnamefont{Lai}}, \bibinfo
  {author} {\bibfnamefont{M.}~\bibnamefont{Guzzi}}, \bibinfo {author}
  {\bibfnamefont{J.}~\bibnamefont{Huston}}, \bibinfo {author}
  {\bibfnamefont{Z.}~\bibnamefont{Li}}, \bibinfo {author}
  {\bibfnamefont{P.~M.}\ \bibnamefont{Nadolsky}}, \emph{et~al.},\ }%
  \bibfield{journal}{%
  \Doi{10.1103/PhysRevD.82.074024}{\bibinfo {journal} {Phys.Rev.}}\ }%
  \textbf{\bibinfo {volume} {D82}},\ \bibinfo {pages} {074024} (\bibinfo {year}
  {2010}),\ \Eprint{http://arxiv.org/abs/1007.2241}{arXiv:1007.2241 [hep-ph]}%
  \bibAnnoteFile{NoStop}{Lai:2010vv}%
\bibitem{Aad:2014cka}%
  \BibitemOpen
  \bibfield{author}{%
  \bibinfo {author} {\bibfnamefont{G.}~\bibnamefont{Aad}} \emph{et~al.}
  (\bibinfo {collaboration} {ATLAS Collaboration}),\ }%
  \bibfield{journal}{%
  \Doi{10.1103/PhysRevD.90.052005}{\bibinfo {journal} {Phys.Rev.}}\ }%
  \textbf{\bibinfo {volume} {D90}},\ \bibinfo {pages} {052005} (\bibinfo {year}
  {2014}),\ \Eprint{http://arxiv.org/abs/1405.4123}{arXiv:1405.4123 [hep-ex]}%
  \bibAnnoteFile{NoStop}{Aad:2014cka}%
\bibitem{Godfrey:2013eta}%
  \BibitemOpen
  \bibfield{author}{%
  \bibinfo {author} {\bibfnamefont{S.}~\bibnamefont{Godfrey}}\ and\ \bibinfo
  {author} {\bibfnamefont{T.}~\bibnamefont{Martin}}}%
   (\bibinfo {year} {2013}),\
  \Eprint{http://arxiv.org/abs/1309.1688}{arXiv:1309.1688 [hep-ph]}%
  \bibAnnoteFile{NoStop}{Godfrey:2013eta}%
\bibitem{Rizzo:2014xma}%
  \BibitemOpen
  \bibfield{author}{%
  \bibinfo {author} {\bibfnamefont{T.~G.}\ \bibnamefont{Rizzo}},\ }%
  \bibfield{journal}{%
  \Doi{10.1103/PhysRevD.89.095022}{\bibinfo {journal} {Phys.Rev.}}\ }%
  \textbf{\bibinfo {volume} {D89}},\ \bibinfo {pages} {095022} (\bibinfo {year}
  {2014}),\ \Eprint{http://arxiv.org/abs/1403.5465}{arXiv:1403.5465 [hep-ph]}%
  \bibAnnoteFile{NoStop}{Rizzo:2014xma}%
\bibitem{Diaz:2004fs}%
  \BibitemOpen
  \bibfield{author}{%
  \bibinfo {author} {\bibfnamefont{R.~A.}\ \bibnamefont{Diaz}}, \bibinfo
  {author} {\bibfnamefont{R.}~\bibnamefont{Martinez}},\ and\ \bibinfo {author}
  {\bibfnamefont{F.}~\bibnamefont{Ochoa}},\ }%
  \bibfield{journal}{%
  \Doi{10.1103/PhysRevD.72.035018}{\bibinfo {journal} {Phys.Rev.}}\ }%
  \textbf{\bibinfo {volume} {D72}},\ \bibinfo {pages} {035018} (\bibinfo {year}
  {2005}),\ \Eprint{http://arxiv.org/abs/hep-ph/0411263}{arXiv:hep-ph/0411263
  [hep-ph]}%
  \bibAnnoteFile{NoStop}{Diaz:2004fs}%
\bibitem{CarcamoHernandez:2005ka}%
  \BibitemOpen
  \bibfield{author}{%
  \bibinfo {author} {\bibfnamefont{A.}~\bibnamefont{Carcamo~Hernandez}},
  \bibinfo {author} {\bibfnamefont{R.}~\bibnamefont{Martinez}},\ and\ \bibinfo
  {author} {\bibfnamefont{F.}~\bibnamefont{Ochoa}},\ }%
  \bibfield{journal}{%
  \Doi{10.1103/PhysRevD.73.035007}{\bibinfo {journal} {Phys.Rev.}}\ }%
  \textbf{\bibinfo {volume} {D73}},\ \bibinfo {pages} {035007} (\bibinfo {year}
  {2006}),\ \Eprint{http://arxiv.org/abs/hep-ph/0510421}{arXiv:hep-ph/0510421
  [hep-ph]}%
  \bibAnnoteFile{NoStop}{CarcamoHernandez:2005ka}%
\bibitem{Pleitez:1993gc}%
  \BibitemOpen
  \bibfield{author}{%
  \bibinfo {author} {\bibfnamefont{V.}~\bibnamefont{Pleitez}}\ and\ \bibinfo
  {author} {\bibfnamefont{M.}~\bibnamefont{Tonasse}},\ }%
  \bibfield{journal}{%
  \Doi{10.1103/PhysRevD.48.5274}{\bibinfo {journal} {Phys.Rev.}}\ }%
  \textbf{\bibinfo {volume} {D48}},\ \bibinfo {pages} {5274} (\bibinfo {year}
  {1993}),\ \Eprint{http://arxiv.org/abs/hep-ph/9302201}{arXiv:hep-ph/9302201
  [hep-ph]}%
  \bibAnnoteFile{NoStop}{Pleitez:1993gc}%
\bibitem{Ng:1992st}%
  \BibitemOpen
  \bibfield{author}{%
  \bibinfo {author} {\bibfnamefont{D.}~\bibnamefont{Ng}},\ }%
  \bibfield{journal}{%
  \Doi{10.1103/PhysRevD.49.4805}{\bibinfo {journal} {Phys.Rev.}}\ }%
  \textbf{\bibinfo {volume} {D49}},\ \bibinfo {pages} {4805} (\bibinfo {year}
  {1994}),\ \Eprint{http://arxiv.org/abs/hep-ph/9212284}{arXiv:hep-ph/9212284
  [hep-ph]}%
  \bibAnnoteFile{NoStop}{Ng:1992st}%
\bibitem{Epele:1995vv}%
  \BibitemOpen
  \bibfield{author}{%
  \bibinfo {author} {\bibfnamefont{L.}~\bibnamefont{Epele}}, \bibinfo {author}
  {\bibfnamefont{H.}~\bibnamefont{Fanchiotti}}, \bibinfo {author}
  {\bibfnamefont{C.}~\bibnamefont{Garcia~Canal}},\ and\ \bibinfo {author}
  {\bibfnamefont{D.}~\bibnamefont{Gomez~Dumm}},\ }%
  \bibfield{journal}{%
  \Doi{10.1016/0370-2693(94)01415-9}{\bibinfo {journal} {Phys.Lett.}}\ }%
  \textbf{\bibinfo {volume} {B343}},\ \bibinfo {pages} {291} (\bibinfo {year}
  {1995})%
  \bibAnnoteFile{NoStop}{Epele:1995vv}%
\bibitem{Gutierrez:2004sba}%
  \BibitemOpen
  \bibfield{author}{%
  \bibinfo {author} {\bibfnamefont{D.~A.}\ \bibnamefont{Gutierrez}}, \bibinfo
  {author} {\bibfnamefont{W.~A.}\ \bibnamefont{Ponce}},\ and\ \bibinfo {author}
  {\bibfnamefont{L.~A.}\ \bibnamefont{Sanchez}},\ }%
  \bibfield{journal}{%
  \Doi{10.1140/epjc/s2006-02513-y}{\bibinfo {journal} {Eur.Phys.J.}}\ }%
  \textbf{\bibinfo {volume} {C46}},\ \bibinfo {pages} {497} (\bibinfo {year}
  {2006}),\ \Eprint{http://arxiv.org/abs/hep-ph/0411077}{arXiv:hep-ph/0411077
  [hep-ph]}%
  \bibAnnoteFile{NoStop}{Gutierrez:2004sba}%
\bibitem{Ponce:2006au}%
  \BibitemOpen
  \bibfield{author}{%
  \bibinfo {author} {\bibfnamefont{W.~A.}\ \bibnamefont{Ponce}}\ and\ \bibinfo
  {author} {\bibfnamefont{O.}~\bibnamefont{Zapata}},\ }%
  \bibfield{journal}{%
  \Doi{10.1103/PhysRevD.74.093007}{\bibinfo {journal} {Phys.Rev.}}\ }%
  \textbf{\bibinfo {volume} {D74}},\ \bibinfo {pages} {093007} (\bibinfo {year}
  {2006}),\ \Eprint{http://arxiv.org/abs/hep-ph/0611082}{arXiv:hep-ph/0611082
  [hep-ph]}%
  \bibAnnoteFile{NoStop}{Ponce:2006au}%
\bibitem{Anderson:2005ab}%
  \BibitemOpen
  \bibfield{author}{%
  \bibinfo {author} {\bibfnamefont{D.~L.}\ \bibnamefont{Anderson}}\ and\
  \bibinfo {author} {\bibfnamefont{M.}~\bibnamefont{Sher}},\ }%
  \bibfield{journal}{%
  \Doi{10.1103/PhysRevD.72.095014}{\bibinfo {journal} {Phys.Rev.}}\ }%
  \textbf{\bibinfo {volume} {D72}},\ \bibinfo {pages} {095014} (\bibinfo {year}
  {2005}),\ \Eprint{http://arxiv.org/abs/hep-ph/0509200}{arXiv:hep-ph/0509200
  [hep-ph]}%
  \bibAnnoteFile{NoStop}{Anderson:2005ab}%
\bibitem{Sanchez:2001ua}%
  \BibitemOpen
  \bibfield{author}{%
  \bibinfo {author} {\bibfnamefont{L.~A.}\ \bibnamefont{Sanchez}}, \bibinfo
  {author} {\bibfnamefont{W.~A.}\ \bibnamefont{Ponce}},\ and\ \bibinfo {author}
  {\bibfnamefont{R.}~\bibnamefont{Martinez}},\ }%
  \bibfield{journal}{%
  \Doi{10.1103/PhysRevD.64.075013}{\bibinfo {journal} {Phys.Rev.}}\ }%
  \textbf{\bibinfo {volume} {D64}},\ \bibinfo {pages} {075013} (\bibinfo {year}
  {2001}),\ \Eprint{http://arxiv.org/abs/hep-ph/0103244}{arXiv:hep-ph/0103244
  [hep-ph]}%
  \bibAnnoteFile{NoStop}{Sanchez:2001ua}%
\bibitem{Martinez:2001mu}%
  \BibitemOpen
  \bibfield{author}{%
  \bibinfo {author} {\bibfnamefont{R.}~\bibnamefont{Martinez}}, \bibinfo
  {author} {\bibfnamefont{W.~A.}\ \bibnamefont{Ponce}},\ and\ \bibinfo {author}
  {\bibfnamefont{L.~A.}\ \bibnamefont{Sanchez}},\ }%
  \bibfield{journal}{%
  \Doi{10.1103/PhysRevD.65.055013}{\bibinfo {journal} {Phys.Rev.}}\ }%
  \textbf{\bibinfo {volume} {D65}},\ \bibinfo {pages} {055013} (\bibinfo {year}
  {2002}),\ \Eprint{http://arxiv.org/abs/hep-ph/0110246}{arXiv:hep-ph/0110246
  [hep-ph]}%
  \bibAnnoteFile{NoStop}{Martinez:2001mu}%
\bibitem{Witten:1985xc}%
  \BibitemOpen
  \bibfield{author}{%
  \bibinfo {author} {\bibfnamefont{E.}~\bibnamefont{Witten}},\ }%
  \bibfield{journal}{%
  \Doi{10.1016/0550-3213(85)90603-0}{\bibinfo {journal} {Nucl.Phys.}}\ }%
  \textbf{\bibinfo {volume} {B258}},\ \bibinfo {pages} {75} (\bibinfo {year}
  {1985})%
  \bibAnnoteFile{NoStop}{Witten:1985xc}%
\bibitem{Pati:1974yy}%
  \BibitemOpen
  \bibfield{author}{%
  \bibinfo {author} {\bibfnamefont{J.~C.}\ \bibnamefont{Pati}}\ and\ \bibinfo
  {author} {\bibfnamefont{A.}~\bibnamefont{Salam}},\ }%
  \bibfield{journal}{%
  \Doi{10.1103/PhysRevD.10.275, 10.1103/PhysRevD.11.703.2}{\bibinfo {journal}
  {Phys.Rev.}}\ }%
  \textbf{\bibinfo {volume} {D10}},\ \bibinfo {pages} {275} (\bibinfo {year}
  {1974})%
  \bibAnnoteFile{NoStop}{Pati:1974yy}%
\bibitem{Mohapatra:1974hk}%
  \BibitemOpen
  \bibfield{author}{%
  \bibinfo {author} {\bibfnamefont{R.~N.}\ \bibnamefont{Mohapatra}}\ and\
  \bibinfo {author} {\bibfnamefont{J.~C.}\ \bibnamefont{Pati}},\ }%
  \bibfield{journal}{%
  \Doi{10.1103/PhysRevD.11.566}{\bibinfo {journal} {Phys.Rev.}}\ }%
  \textbf{\bibinfo {volume} {D11}},\ \bibinfo {pages} {566} (\bibinfo {year}
  {1975})%
  \bibAnnoteFile{NoStop}{Mohapatra:1974hk}%
\bibitem{Mohapatra:1974gc}%
  \BibitemOpen
  \bibfield{author}{%
  \bibinfo {author} {\bibfnamefont{R.}~\bibnamefont{Mohapatra}}\ and\ \bibinfo
  {author} {\bibfnamefont{J.~C.}\ \bibnamefont{Pati}},\ }%
  \bibfield{journal}{%
  \Doi{10.1103/PhysRevD.11.2558}{\bibinfo {journal} {Phys.Rev.}}\ }%
  \textbf{\bibinfo {volume} {D11}},\ \bibinfo {pages} {2558} (\bibinfo {year}
  {1975})%
  \bibAnnoteFile{NoStop}{Mohapatra:1974gc}%
\bibitem{Ma:1995xk}%
  \BibitemOpen
  \bibfield{author}{%
  \bibinfo {author} {\bibfnamefont{E.}~\bibnamefont{Ma}},\ }%
  \bibfield{journal}{%
  \Doi{10.1016/0370-2693(96)00524-2}{\bibinfo {journal} {Phys.Lett.}}\ }%
  \textbf{\bibinfo {volume} {B380}},\ \bibinfo {pages} {286} (\bibinfo {year}
  {1996}),\ \Eprint{http://arxiv.org/abs/hep-ph/9507348}{arXiv:hep-ph/9507348
  [hep-ph]}%
  \bibAnnoteFile{NoStop}{Ma:1995xk}%
\bibitem{King:2005jy}%
  \BibitemOpen
  \bibfield{author}{%
  \bibinfo {author} {\bibfnamefont{S.}~\bibnamefont{King}}, \bibinfo {author}
  {\bibfnamefont{S.}~\bibnamefont{Moretti}},\ and\ \bibinfo {author}
  {\bibfnamefont{R.}~\bibnamefont{Nevzorov}},\ }%
  \bibfield{journal}{%
  \Doi{10.1103/PhysRevD.73.035009}{\bibinfo {journal} {Phys.Rev.}}\ }%
  \textbf{\bibinfo {volume} {D73}},\ \bibinfo {pages} {035009} (\bibinfo {year}
  {2006}),\ \Eprint{http://arxiv.org/abs/hep-ph/0510419}{arXiv:hep-ph/0510419
  [hep-ph]}%
  \bibAnnoteFile{NoStop}{King:2005jy}%
\bibitem{Erler:2002pr}%
  \BibitemOpen
  \bibfield{author}{%
  \bibinfo {author} {\bibfnamefont{J.}~\bibnamefont{Erler}}, \bibinfo {author}
  {\bibfnamefont{P.}~\bibnamefont{Langacker}},\ and\ \bibinfo {author}
  {\bibfnamefont{T.-j.}\ \bibnamefont{Li}},\ }%
  \bibfield{journal}{%
  \Doi{10.1103/PhysRevD.66.015002}{\bibinfo {journal} {Phys.Rev.}}\ }%
  \textbf{\bibinfo {volume} {D66}},\ \bibinfo {pages} {015002} (\bibinfo {year}
  {2002}),\ \Eprint{http://arxiv.org/abs/hep-ph/0205001}{arXiv:hep-ph/0205001
  [hep-ph]}%
  \bibAnnoteFile{NoStop}{Erler:2002pr}%
\bibitem{Kang:2004pp}%
  \BibitemOpen
  \bibfield{author}{%
  \bibinfo {author} {\bibfnamefont{J.}~\bibnamefont{Kang}}, \bibinfo {author}
  {\bibfnamefont{P.}~\bibnamefont{Langacker}}, \bibinfo {author}
  {\bibfnamefont{T.-j.}\ \bibnamefont{Li}},\ and\ \bibinfo {author}
  {\bibfnamefont{T.}~\bibnamefont{Liu}},\ }%
  \bibfield{journal}{%
  \Doi{10.1103/PhysRevLett.94.061801}{\bibinfo {journal} {Phys.Rev.Lett.}}\ }%
  \textbf{\bibinfo {volume} {94}},\ \bibinfo {pages} {061801} (\bibinfo {year}
  {2005}),\ \Eprint{http://arxiv.org/abs/hep-ph/0402086}{arXiv:hep-ph/0402086
  [hep-ph]}%
  \bibAnnoteFile{NoStop}{Kang:2004pp}%
\bibitem{Erler:2009jh}%
  \BibitemOpen
  \bibfield{author}{%
  \bibinfo {author} {\bibfnamefont{J.}~\bibnamefont{Erler}}, \bibinfo {author}
  {\bibfnamefont{P.}~\bibnamefont{Langacker}}, \bibinfo {author}
  {\bibfnamefont{S.}~\bibnamefont{Munir}},\ and\ \bibinfo {author}
  {\bibfnamefont{E.}~\bibnamefont{Rojas}},\ }%
  \bibfield{journal}{%
  \Doi{10.1088/1126-6708/2009/08/017}{\bibinfo {journal} {JHEP}}\ }%
  \textbf{\bibinfo {volume} {0908}},\ \bibinfo {pages} {017} (\bibinfo {year}
  {2009}),\ \Eprint{http://arxiv.org/abs/0906.2435}{arXiv:0906.2435 [hep-ph]}%
  \bibAnnoteFile{NoStop}{Erler:2009jh}%
\bibitem{Erler:2009ut}%
  \BibitemOpen
  \bibfield{author}{%
  \bibinfo {author} {\bibfnamefont{J.}~\bibnamefont{Erler}}, \bibinfo {author}
  {\bibfnamefont{P.}~\bibnamefont{Langacker}}, \bibinfo {author}
  {\bibfnamefont{S.}~\bibnamefont{Munir}},\ and\ \bibinfo {author}
  {\bibfnamefont{E.}~\bibnamefont{Rojas}},\ }%
  \bibfield{journal}{%
  \Doi{10.1063/1.3327731}{\bibinfo {journal} {AIP Conf.Proc.}}\ }%
  \textbf{\bibinfo {volume} {1200}},\ \bibinfo {pages} {790} (\bibinfo {year}
  {2010}),\ \Eprint{http://arxiv.org/abs/0910.0269}{arXiv:0910.0269 [hep-ph]}%
  \bibAnnoteFile{NoStop}{Erler:2009ut}%
\bibitem{delAguila:2010mx}%
  \BibitemOpen
  \bibfield{author}{%
  \bibinfo {author} {\bibfnamefont{F.}~\bibnamefont{del Aguila}}, \bibinfo
  {author} {\bibfnamefont{J.}~\bibnamefont{de~Blas}},\ and\ \bibinfo {author}
  {\bibfnamefont{M.}~\bibnamefont{Perez-Victoria}},\ }%
  \bibfield{journal}{%
  \Doi{10.1007/JHEP09(2010)033}{\bibinfo {journal} {JHEP}}\ }%
  \textbf{\bibinfo {volume} {1009}},\ \bibinfo {pages} {033} (\bibinfo {year}
  {2010}),\ \Eprint{http://arxiv.org/abs/1005.3998}{arXiv:1005.3998 [hep-ph]}%
  \bibAnnoteFile{NoStop}{delAguila:2010mx}%
\bibitem{Erler:2010uy}%
  \BibitemOpen
  \bibfield{author}{%
  \bibinfo {author} {\bibfnamefont{J.}~\bibnamefont{Erler}}, \bibinfo {author}
  {\bibfnamefont{P.}~\bibnamefont{Langacker}}, \bibinfo {author}
  {\bibfnamefont{S.}~\bibnamefont{Munir}},\ and\ \bibinfo {author}
  {\bibfnamefont{E.}~\bibnamefont{rojas}}}%
   (\bibinfo {year} {2010}),\
  \Eprint{http://arxiv.org/abs/1010.3097}{arXiv:1010.3097 [hep-ph]}%
  \bibAnnoteFile{NoStop}{Erler:2010uy}%
\bibitem{Agashe:2014kda}%
  \BibitemOpen
  \bibfield{author}{%
  \bibinfo {author} {\bibfnamefont{K.}~\bibnamefont{Olive}} \emph{et~al.}
  (\bibinfo {collaboration} {Particle Data Group}),\ }%
  \bibfield{journal}{%
  \Doi{10.1088/1674-1137/38/9/090001}{\bibinfo {journal} {Chin.Phys.}}\ }%
  \textbf{\bibinfo {volume} {C38}},\ \bibinfo {pages} {090001} (\bibinfo {year}
  {2014})%
  \bibAnnoteFile{NoStop}{Agashe:2014kda}%
\bibitem{Aad:2012hf}%
  \BibitemOpen
  \bibfield{author}{%
  \bibinfo {author} {\bibfnamefont{G.}~\bibnamefont{Aad}} \emph{et~al.}
  (\bibinfo {collaboration} {ATLAS Collaboration}),\ }%
  \bibfield{journal}{%
  \Doi{10.1007/JHEP11(2012)138}{\bibinfo {journal} {JHEP}}\ }%
  \textbf{\bibinfo {volume} {1211}},\ \bibinfo {pages} {138} (\bibinfo {year}
  {2012}),\ \Eprint{http://arxiv.org/abs/1209.2535}{arXiv:1209.2535 [hep-ex]}%
  \bibAnnoteFile{NoStop}{Aad:2012hf}%
\bibitem{Ellis:1991qj}%
  \BibitemOpen
  \bibfield{author}{%
  \bibinfo {author} {\bibfnamefont{R.~K.}\ \bibnamefont{Ellis}}, \bibinfo
  {author} {\bibfnamefont{W.~J.}\ \bibnamefont{Stirling}},\ and\ \bibinfo
  {author} {\bibfnamefont{B.}~\bibnamefont{Webber}},\ }%
  \bibfield{journal}{%
  \bibinfo {journal} {Camb.Monogr.Part.Phys.Nucl.Phys.Cosmol.}\ }%
  \textbf{\bibinfo {volume} {8}},\ \bibinfo {pages} {1} (\bibinfo {year}
  {1996})%
  \bibAnnoteFile{NoStop}{Ellis:1991qj}%
\bibitem{Kang:2004bz}%
  \BibitemOpen
  \bibfield{author}{%
  \bibinfo {author} {\bibfnamefont{J.}~\bibnamefont{Kang}}\ and\ \bibinfo
  {author} {\bibfnamefont{P.}~\bibnamefont{Langacker}},\ }%
  \bibfield{journal}{%
  \Doi{10.1103/PhysRevD.71.035014}{\bibinfo {journal} {Phys.Rev.}}\ }%
  \textbf{\bibinfo {volume} {D71}},\ \bibinfo {pages} {035014} (\bibinfo {year}
  {2005}),\ \Eprint{http://arxiv.org/abs/hep-ph/0412190}{arXiv:hep-ph/0412190
  [hep-ph]}%
  \bibAnnoteFile{NoStop}{Kang:2004bz}%
\bibitem{Baur:2001ze}%
  \BibitemOpen
  \bibfield{author}{%
  \bibinfo {author} {\bibfnamefont{U.}~\bibnamefont{Baur}}, \bibinfo {author}
  {\bibfnamefont{O.}~\bibnamefont{Brein}}, \bibinfo {author}
  {\bibfnamefont{W.}~\bibnamefont{Hollik}}, \bibinfo {author}
  {\bibfnamefont{C.}~\bibnamefont{Schappacher}},\ and\ \bibinfo {author}
  {\bibfnamefont{D.}~\bibnamefont{Wackeroth}},\ }%
  \bibfield{journal}{%
  \Doi{10.1103/PhysRevD.65.033007}{\bibinfo {journal} {Phys.Rev.}}\ }%
  \textbf{\bibinfo {volume} {D65}},\ \bibinfo {pages} {033007} (\bibinfo {year}
  {2002}),\ \Eprint{http://arxiv.org/abs/hep-ph/0108274}{arXiv:hep-ph/0108274
  [hep-ph]}%
  \bibAnnoteFile{NoStop}{Baur:2001ze}%
\bibitem{Dias:2005xj}%
  \BibitemOpen
  \bibfield{author}{%
  \bibinfo {author} {\bibfnamefont{A.~G.}\ \bibnamefont{Dias}}, \bibinfo
  {author} {\bibfnamefont{J.}~\bibnamefont{Montero}},\ and\ \bibinfo {author}
  {\bibfnamefont{V.}~\bibnamefont{Pleitez}},\ }%
  \bibfield{journal}{%
  \Doi{10.1016/j.physletb.2006.04.015}{\bibinfo {journal} {Phys.Lett.}}\ }%
  \textbf{\bibinfo {volume} {B637}},\ \bibinfo {pages} {85} (\bibinfo {year}
  {2006}),\ \Eprint{http://arxiv.org/abs/hep-ph/0511084}{arXiv:hep-ph/0511084
  [hep-ph]}%
  \bibAnnoteFile{NoStop}{Dias:2005xj}%
\end{thebibliography}%






\end{document}